\documentclass[journal, compsoc]{IEEEtran}

\usepackage{cite}
\usepackage{multirow}
\usepackage[pdftex]{graphicx}
\usepackage{amsmath}
\usepackage{bm}
\usepackage{algorithmic}
\usepackage{array}
\usepackage{booktabs}
\usepackage{hyperref}
\usepackage{bbding}
\usepackage{pifont}
\usepackage{caption}
\usepackage{amsfonts}
\usepackage{booktabs}
\usepackage{hyperref}
\usepackage{graphicx} 
\usepackage{pifont}
\usepackage{enumitem}
\usepackage{mydef}
\usepackage{threeparttable} 
\usepackage[capitalize]{cleveref}
\crefname{figure}{Figure}{Figures}
\crefname{table}{Table}{Tables}
\crefformat{equation}{Eq.~(#2#1#3)}

\newcommand{\tabref}[1]{Table~\ref{#1}}

\usepackage{tikz}
\usepackage[disable]{todonotes}
\usetikzlibrary{fadings}
\usetikzlibrary{mindmap,trees}
\usetikzlibrary{arrows,automata,shapes,positioning,shadows,trees}
\usetikzlibrary{shapes,snakes,shadows}
\usetikzlibrary{shapes.arrows}
\usetikzlibrary{calc,shapes, positioning}
\usetikzlibrary{decorations.text}
\usetikzlibrary{matrix,chains,positioning,decorations.pathreplacing,arrows}
\usetikzlibrary{bayesnet}
\usepackage[edges]{forest}
\usetikzlibrary{shadows.blur}
\usetikzlibrary{shapes.geometric}

\begin{document}

\title{Materials Generation in the Era of Artificial Intelligence: A Comprehensive Survey}

\author{Zhixun Li$^{\dagger}$, Bin Cao$^{\dagger}$, Rui Jiao$^{\dagger}$, Liang Wang$^{\dagger}$, Ding Wang, Yang Liu, Dingshuo Chen, \\ Jia Li, Qiang Liu, Yu Rong, Liang Wang, Tong-yi Zhang, Jeffrey Xu Yu

\IEEEcompsocitemizethanks{\IEEEcompsocthanksitem $\dagger$ The first four authors contributed equally to this work.

\IEEEcompsocthanksitem Zhixun Li, Jeffrey Xu Yu: The Chinese University of Hong Kong. \{zxli, yu\}@se.cuhk.edu.hk

\IEEEcompsocthanksitem Bin Cao, Yang Liu, Jia Li: The Hong Kong University of Science and Technology (Guangzhou). bcao686@connect.hkust-gz.edu.cn, yliukj@connect.ust.hk, jialee@ust.hk

\IEEEcompsocthanksitem Rui Jiao: Tsinghua University. jiaor21@mails.tsinghua.edu.cn

\IEEEcompsocthanksitem Liang Wang, Ding Wang, Dingshuo Chen, Qiang Liu, Liang Wang: Institute of Automation, Chinese Academy of Sciences. \{liang.wang, dingshuo.chen\}@cripac.ia.ac.cn, wangding2024@ia.ac.cn, \{qiang.liu, wangliang\}@nlpr.ia.ac.cn

\IEEEcompsocthanksitem Yu Rong: DAMO Academy, Alibaba Group. yu.rong@hotmail.com

\IEEEcompsocthanksitem Tong-yi Zhang: The Hong Kong University of Science and Technology (Guangzhou) and Shanghai University. mezhangt@hkust-gz.edu.cn
}}

\markboth{Journal of \LaTeX\ Class Files,~Vol.~14, No.~8, August~2015}%
{Shell \MakeLowercase{\textit{et al.}}: Bare Demo of IEEEtran.cls for IEEE Journals}

\maketitle

\begin{abstract}

Materials are the foundation of modern society, underpinning advancements in energy, electronics, healthcare, transportation, and infrastructure. The ability to discover and design new materials with tailored properties is critical to solving some of the most pressing global challenges. In recent years, the growing availability of high-quality materials data combined with rapid advances in Artificial Intelligence (AI) has opened new opportunities for accelerating materials discovery. Data-driven generative models provide a powerful tool for materials design by directly create novel materials that satisfy predefined property requirements. Despite the proliferation of related work, there remains a notable lack of up-to-date and systematic surveys in this area. To fill this gap, this paper provides a comprehensive overview of recent progress in AI-driven materials generation. We first organize various types of materials and illustrate multiple representations of crystalline materials. We then provide a detailed summary and taxonomy of current AI-driven materials generation approaches. Furthermore, we discuss the common evaluation metrics and summarize open-source codes and benchmark datasets. Finally, we conclude with potential future directions and challenges in this fast-growing field. The related sources can be found at \href{https://github.com/ZhixunLEE/Awesome-AI-for-Materials-Generation}{https://github.com/ZhixunLEE/Awesome-AI-for-Materials-Generation}.

\end{abstract}

\begin{IEEEkeywords}
AI for Science, Materials Discovery, Generative Models, Equivariance, Invariance.
\end{IEEEkeywords}

\IEEEpeerreviewmaketitle

\section{Introduction}

Materials discovery is a fundamental driver of technological advancement with direct impact on real-world challenges. From energy systems and electronics to biomedical devices and sustainable manufacturing, novel materials enable new functionalities and improved performance \cite{ramprasad2017machine,meredig2014combinatorial,oliynyk2016high,cao2024active,raccuglia2016machine,ward2016general,luo2021graphdf,li2025machine,chanussot2021open,cao2022domain}. As global demands grow, the rational design and exploration of advanced materials are critical for enabling scalable, efficient, and sustainable solutions.

Recent advances such as high-throughput screening \cite{curtarolo2013high}, open-access materials databases \cite{gravzulis2012crystallography,jain2020materials,curtarolo2012aflow,barroso2024open,draxl2019nomad,caobin2025hkustcrystdb}, and machine learning-based property prediction \cite{cao2025asugnn,chen2019graph,unke2021machine} have significantly accelerated the materials discovery process. However, identifying materials with optimal performance remains a major challenge. Small changes in atomic structure or composition can result in substantial variations in material properties, rendering the search space highly complex and non-linear \cite{cao2025asugnn}. This complexity is closely linked to the nature of the potential energy surface governing lattice structures \cite{bloch2022strongly}.

\begin{figure*}[h!]
\centering
\includegraphics[width=\textwidth]{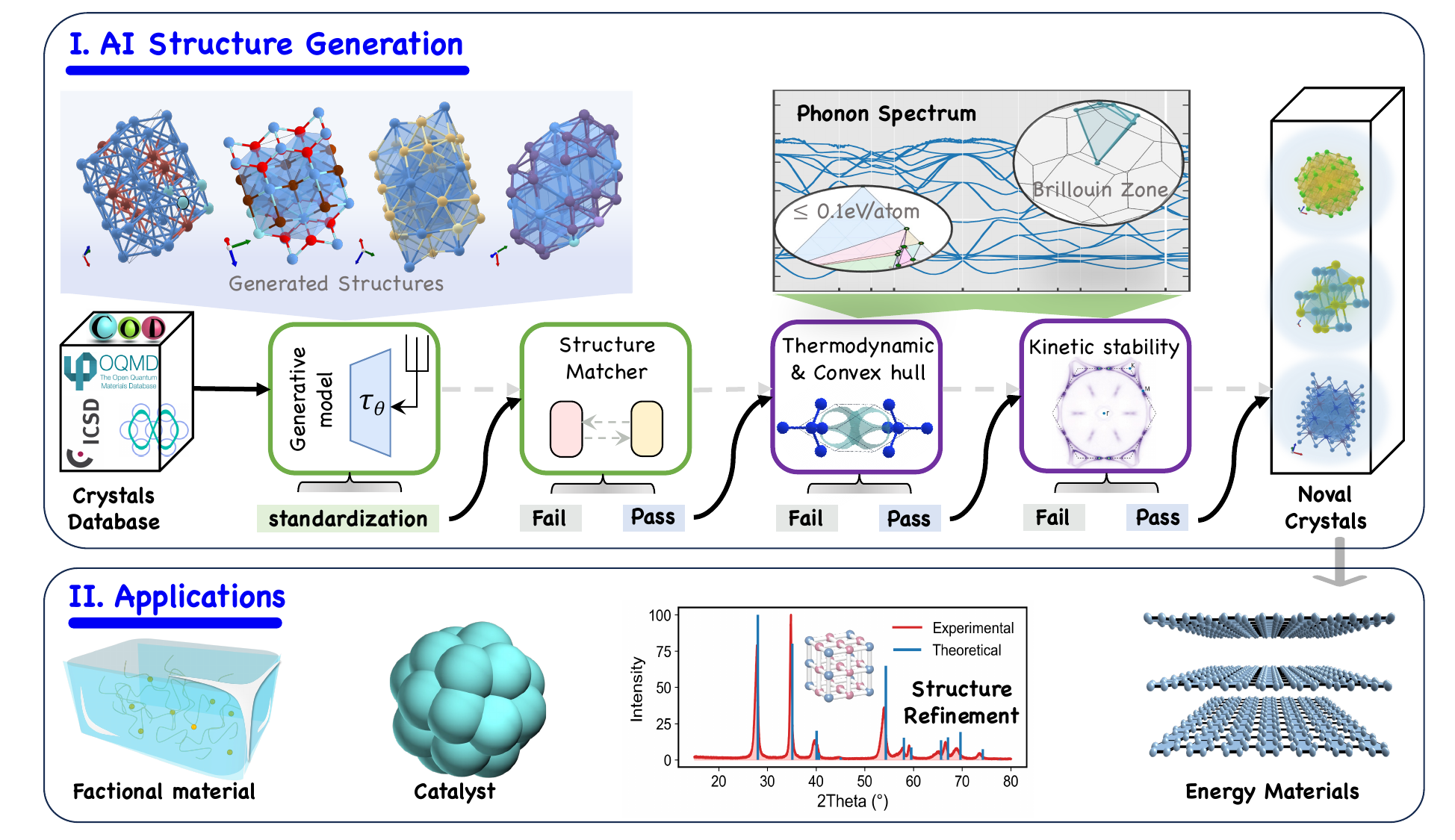}
\caption{A theoretically novel crystal structure should be standardized, unique, and pass both thermodynamic and kinetic stability tests. Such a structure holds the potential to revolutionize society by enabling the development of new functional materials, catalysts and energy-related materials. The generated structure provides an initial topology for diffraction-based techniques to confirm unknown crystal structures.}
\label{fig:crystgen}
\end{figure*}

To address these challenges, there is growing interest in the inverse design of materials. For example, the launch of the USPEX (Universal Structure Predictor: Evolutionary Xtallography) project marked a major milestone in non-empirical crystal structure prediction \cite{glass2006uspex}. Inverse design aims to directly generate material structures that satisfy specific — and often rare or conflicting — target property constraints. Techniques such as evolutionary algorithms\cite{allahyari2020coevolutionary,zhu2024wycryst,wang2012calypso} and reinforcement learning \cite{law2022upper} have been employed to explore this approach. Promising candidates are then evaluated using density functional theory (DFT) calculations to identify stable and high-performing compounds. However, generating crystal structures remains computationally intensive, particularly when exploring novel materials. This difficulty arises from two primary factors: (1) the intrinsic complexity of representing crystal structures, and (2) the need for robust optimization algorithms capable of efficiently identifying thermodynamically stable configurations within a vast compositional and structural space.

With significant breakthroughs in the three driving forces of AI, computing power, data, and algorithms, over the past decade, material generation has rapidly advanced in the era of AI (Figure \ref{fig:crystgen}). (1) \textbf{Computing power}: In past decade, global computational power has increased by approximately 1,000 times. This growth is largely driven by advances in GPU and AI-specific hardware, alongside the proliferation of cloud computing, enabling the rapid training of large-scale AI models and enhancing computational capabilities across various industries. (2) \textbf{Data}: Increasing number of large datasets from experiments and simulations are available (Table \ref{table:datasetsummary}). For instance, the Material Project database \cite{jain2020materials} includes property calculations for over 60,000 molecules and over 140,000 inorganic compounds. (3) \textbf{Algorithms}: Recent breakthroughs such as large language models \cite{achiam2023gpt,liu2024deepseek,yang2024qwen2} in natural language processing and diffusion models \cite{DDPM,SMLD,ScoreMatching} in computer vision have significantly advanced AI Generated Content (AIGC). These models have redefined content generation by combining scalability, flexibility, and high fidelity across modalities. These three aspects provide the essential fuel for AI-driven materials generation, and none of them can be omitted.

Despite the considerable progress and the proliferation of research community, there remains a notable lack of up-to-date and systematic surveys in this area. The wide variety of material representations, diverse generation methods, and different datasets make it difficult for researchers to gain a comprehensive understanding of the current state of progress. To address this challenge, this paper presents a comprehensive survey of materials generation in the era of artificial intelligence. We first summarize various representations of materials, including geometric graph, textual sequence, SLICES string, and diffraction pattern. We also introduce the different geometric symmetry properties of crystal materials. Next, we categorize the current mainstream generative approaches, including VAE-based, GAN-based, diffusion-based, and autoregressive-based methods, and provide a summary and introduction to the fundamental algorithms as well as the current state-of-the-art models. Finally, we demonstrate the commonly used datasets and evaluation metrics in materials generation research and discuss future directions and challenges.

Our paper seeks to provide a comprehensive review of the materials generation in the era of AI for broader researchers from diverse backgrounds like computer science and materials science. We hope to inspire the research community and have a positive impact on future innovations.

Our contributions are summarized as follows:
\begin{itemize}[leftmargin=*]
    \item \textbf{Detailed Representations of Materials}. We provide a detailed and comprehensive introduction to the representation forms of crystal structures, along with their corresponding mathematical definitions.
    \item \textbf{Systematic Taxonomy of Techniques}. We systematically categorize and compare a large number of existing methods. And we provide a development timeline.
    \item \textbf{Abundant Resources}. We include numerous links to open-source code and datasets to facilitate access and use by readers.
    \item \textbf{Future Directions and Challenges}. We discuss potential future research directions and current challenges that need to be addressed, encouraging further innovation within the research community.
\end{itemize}

\textbf{Roadmap of our survey}. The rest of this survey is organized as follows. Section \ref{sec:related} introduces other related surveys and highlights the differences between our work and theirs. Section \ref{sec:preliminary} provides preliminary definition of materials and generative models. Section \ref{sec:backbone} introduces commonly used encoder architecture. Section \ref{sec:taxonomy} categorizes modern AI-driven materials generation approaches. Section \ref{sec:datasets} delivers available open-access datasets. Section \ref{sec:evaluation} presents evaluation metrics of materials generation. Section \ref{sec:future} provides future research directions and challenges. Finally, Section \ref{sec:conclusion} summarizes the entire paper.

\section{Related Work}
\label{sec:related}

Currently, only a few existing surveys have explored the integration of AI with materials generation. Paper of Chen et al. \cite{chen2025crystal} is the most similar to ours, however, the methods it covers are relatively limited and focus more on techniques. In contrast, our paper provides a comprehensive overview of the material generation field, including data representations, backbones, generative models, datasets, and evaluation metrics. Wang et al. \cite{wang2024crystalline}, Guo et al. \cite{guo2021artificial}, Beyerle et al. \cite{beyerle2023recent}, and Huang et al. \cite{huang2021artificial} systematically analyze the application of AI in material science, covering property prediction, material synthesis, accelerating theoretical computations, and etc. However, none of them provide a detailed survey of materials generation works, nor do they discuss its integration with cutting-edge generative technologies. Our survey offers a comprehensive review of materials generation, providing researchers with a shortcut to quickly grasp the latest advancements in the field.

\section{Preliminary}
\label{sec:preliminary}

In this section, we will provide a clear mathematical definition of materials, including the representation of crystals and symmetry. Additionally, we will introduce the foundational frameworks of current mainstream generative models, including diffusion and autoregressive models. The specific symbol definitions can be found in \tabref{tab:notation}.

\subsection{Definition of Materials}
\subsubsection{Representation of Crystal Structures}

\begin{figure*}[h!]
\centering
\includegraphics[width=\textwidth]{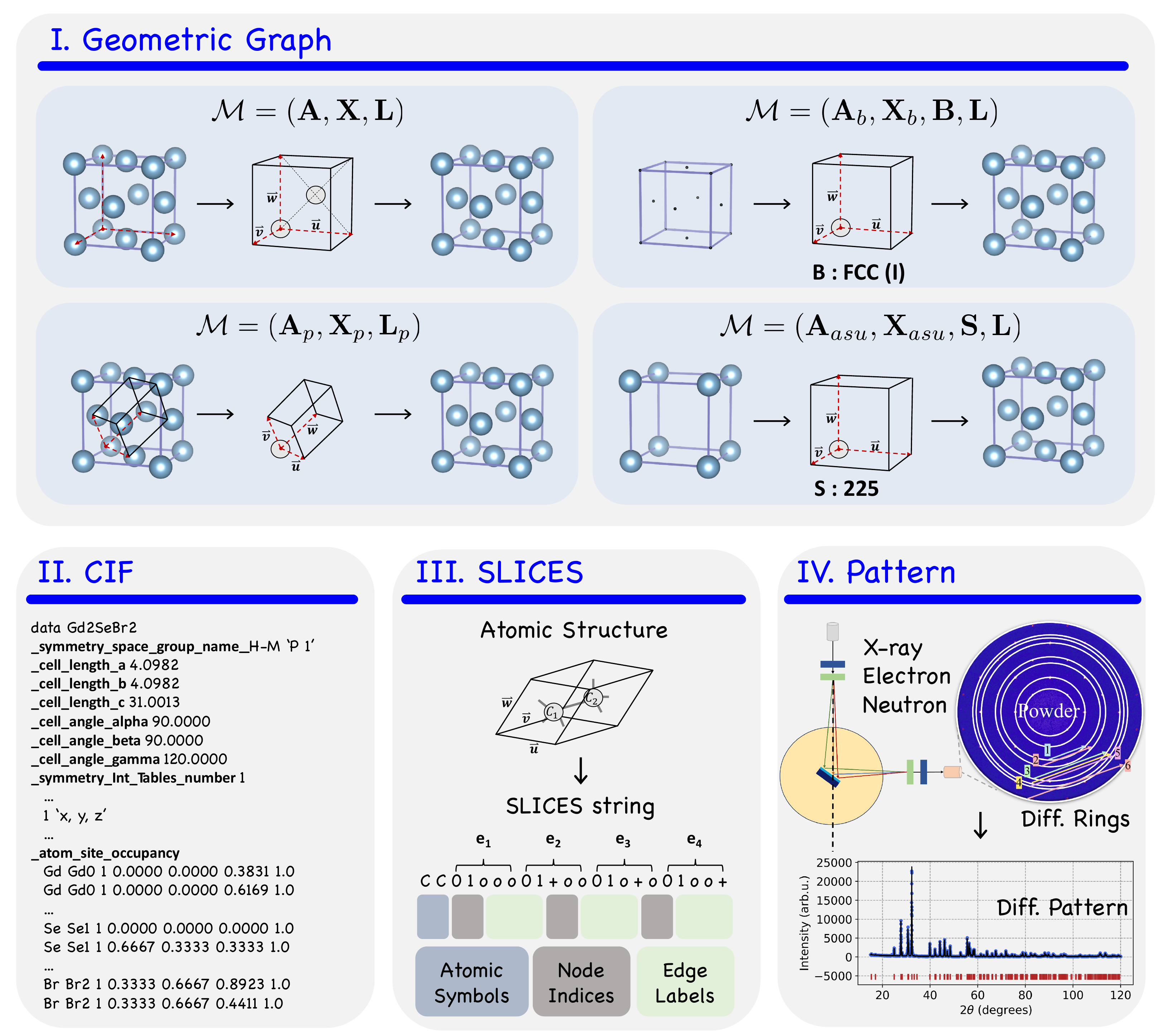}
\caption{Universal crystal structure representation using geometric graphs, textual data (Crystallographic Information File, CIF), SLICES string, and diffraction patterns (X-ray, electron, and neutron).}
\label{fig:cry_rep}
\end{figure*}

\textbf{Geometric Graph Representation}. Solids have both a defined shape and volume. While the particles in a solid can be randomly distributed, they are more likely to form an ordered and repetitive pattern. This ordered arrangement corresponds to a lower energy state compared to a random spatial distribution of strongly interacting atoms or molecules \cite{pecharsky2009fundamentals}. However, not all solids are crystalline or ordered. For instance, glasses possess both shape and volume but exhibit a high degree of disorder, classifying them as amorphous solids. The absence of long-range order generally makes the macroscopic properties of amorphous solids isotropic. In other words, long-range order refers to the regular and, in the simplest case, periodic repetition of atoms or molecules in space. This periodicity enables machine learning models to represent crystals using a chosen periodic unit \cite{cao2025asugnn} (Figure \ref{fig:cry_rep} (I)). 

The periodic structure of an ideal crystal is best described by a unit cell. Knowing the atomic arrangement within a unit allows the entire crystal structure to be reconstructed by propagating the unit along one, two, or three directions.

\textbf{Conventional lattice cell}. The most common periodic unit is the conventional lattice cell, the smallest repeating unit that maximizes symmetry \cite{fjellvaag1994symmetry}. Its geometry is defined by three matrices in 3D space and a set of atomic coordinates:

\begin{equation}
    \mathcal{M}=(\mathbf{A}, \mathbf{X}, \mathbf{L}),
\end{equation}

where $\mathbf{A} = [\mathbf{a}_1, \mathbf{a}_2, \ldots, \mathbf{a}_N] \in \mathbb{R}^{Z \times N}$ be a matrix of feature vectors, where each column vector $\mathbf{a}_i \in \mathbb{R}^Z$ represents the $Z$ attributes of the $i$-th atom in the lattice, $\mathbf{X}=[\mathbf{x}_1,\mathbf{x}_2,\ldots,\mathbf{x}_N]\in \mathbb{R}^{3\times N}$ consists of Cartesian coordinates of atoms, and $\mathbf{L}=[\mathbf{l}_1,\mathbf{l}_2,\mathbf{l_3}]\in\mathbb{R}^{3\times 3}$ represents the edge vectors of the lattice. Propagating the lattice along these vectors in 3D space reconstructs the entire crystal.

\textbf{Bravais grids}. The conventional lattice cell preserves maximum symmetry but is a relatively redundant representation of crystals. If the lattice is further coarse-grained into a Bravais grids, the atom set is compressed. We refer to this as the basis + grid representation.

\begin{equation}  
    \mathcal{M} = (\mathbf{A}_b, \mathbf{X}_b, \mathbf{B},\mathbf{L},),  
\end{equation}  

where $\mathbf{A_{b}} = [\mathbf{a}_1, \mathbf{a}_2, \ldots, \mathbf{a}_{n_1}] \in \mathbb{R}^{Z \times n_1}$ be a matrix of feature vectors (\( n_1 \leq N \)), represents the $Z$ attributes of $n_1$ atoms in the lattice. \( \mathbf{X}_b = [\mathbf{x}_1, \mathbf{x}_2, \dots, \mathbf{x}_{n_1}] \in \mathbb{R}^{3 \times n_1} \) contains the Cartesian coordinates of atoms, and the one-hot vector \( \mathbf{B} \) denotes the Bravais grid type, with 14 possible choices. Generally speaking, the triplet \( (\mathbf{A}_b, \mathbf{X}_b, \mathbf{B}) \) is equivalent to the pair \( (\mathbf{A}, \mathbf{X}) \).

\textbf{Primitive unit cell}. In the extreme case, if the atomic set contains only a single atom (or basis), the smallest representative unit is called the primitive unit cell.  

\begin{equation}  
    \mathcal{M} = (\mathbf{A}_p, \mathbf{X}_p, \mathbf{L}_p),  
\end{equation}

where $\mathbf{A_{p}} = [\mathbf{a}_1, \mathbf{a}_2, \ldots, \mathbf{a}_{n_2}] \in \mathbb{R}^{Z \times n_2}$ be a matrix of feature vectors (\( n_2 \leq n_1 \)), represents the $Z$ attributes of $n_2$ atoms in the lattice. \( \mathbf{X}_p = [\mathbf{x}_1, \mathbf{x}_2, \dots, \mathbf{x}_{n_2}] \in \mathbb{R}^{3 \times n_2} \) contains the Cartesian coordinates of atoms, and \( \mathbf{L}_p = [\mathbf{l}_{p1}, \mathbf{l}_{p2}, \mathbf{l}_{p3}] \in \mathbb{R}^{3 \times 3} \) represents the edge vectors of the primitive unit cell. In principle \( n_2 =1\) when the basis (a cluster atoms) is considered as an abstract point, we define \( n_2\) more generally.

\begin{table}[t]
\caption{Table of notations}
\resizebox{\linewidth}{!}{
\begin{tabular}{ll}
\toprule
\multicolumn{2}{c}{\textbf{Material Notation}} \\ \midrule
  $N \in\mathbb{N}$                & Number of atoms in a conventional unit cell      \\

$n_1\in\mathbb{N}$                & Number of atoms in a Bravais basis       \\

$n_2\in\mathbb{N}$                & Number of atoms in a primitive unit cell      \\

  $n_3\in\mathbb{N}$                & Number of atoms in an asu set.       \\
  
  $\mathcal{M}$                & A crystal structure   \\

  $\mathbf{A},\mathbf{A}_b,\mathbf{A}_p, \mathbf{A}_{asu}$                & Feature attribute vectors of atoms          \\
  
  $\mathbf{X},\mathbf{X}_b,\mathbf{X}_p,\mathbf{X}_{asu} $                & Cartesian atomic coordinates           \\

  $\mathbf{B},\mathbf{S} $                & One-hot type vector of Bravais grid and space group           \\

  $\mathbf{F}\in \mathbb{R}^{3\times n}$                & Fractional atomic coordinates           \\
  $\mathbf{L},\mathbf{L}_p\in\mathbb{R}^{3\times 3}$                & The lattice matrix of conversational, primitive unit cell          \\
  $\mathbf{T}_f\in\mathbb{R}^{4\times 4}$                & The fractional
translation matrix    \\ 
             \bottomrule
\end{tabular}}
\label{tab:notation}
\end{table}

Unlike the conventional lattice cell, the primitive unit cell is the smallest periodic unit, containing only a single atom or basis. However, this size reduction comes at the cost of losing point group symmetry, making it less suitable for crystal representation in machine learning. As a result, the generated structures exhibit degraded symmetry, often collapsing to space group \#1 \cite{binsimxrd}.

\textbf{Asymmetric unit}. To obtain a compact representative set while preserving symmetry in crystal representation, a promising solution is the asymmetric unit (asu) + space group approach:

\begin{equation}  
    \mathcal{M} = (\mathbf{A}_{asu}, \mathbf{X}_{asu}, \mathbf{S}, \mathbf{L}),  
\end{equation}

where $\mathbf{A_{asu}} = [\mathbf{a}_1, \mathbf{a}_2, \ldots, \mathbf{a}_{n_3}] \in \mathbb{R}^{Z \times n_3}$ be a matrix of feature vectors (\( n_3 \leq n_1 \)), represents the $Z$ attributes of $n_3$ atoms in the lattice. \( \mathbf{X}_{asu} = [\mathbf{x}_1, \mathbf{x}_2, \dots, \mathbf{x}_{n_3}] \in \mathbb{R}^{3 \times n_3} \) contains their Cartesian coordinates. The one-hot vector \( \mathbf{S} \) represents the space group type, with 230 possible choices.  Generally speaking, the triplet \( (\mathbf{A}_{asu}, \mathbf{X}_{asu}, \mathbf{S}) \) is equivalent to the pair \( (\mathbf{A}, \mathbf{X}) \).  

\begin{figure*}[h!]
\centering
\includegraphics[width=\textwidth]{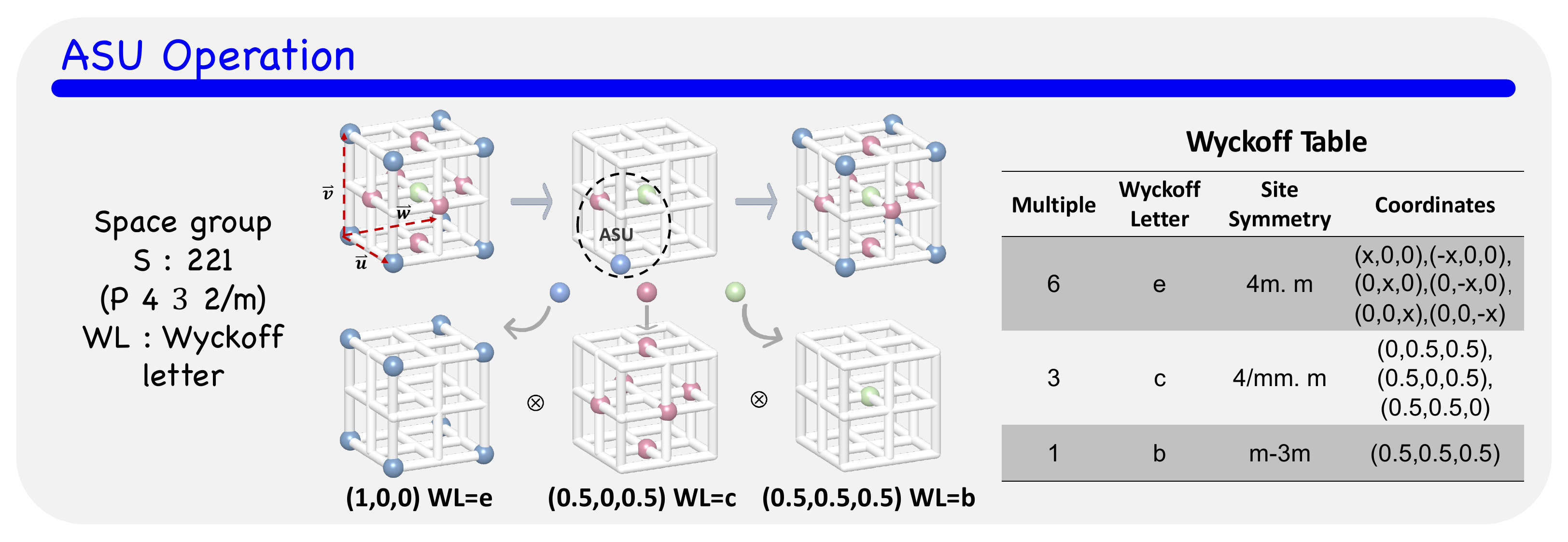}
\caption{The illustration of crystals constructed using the asu and space group.}
\label{fig:asu_show}
\end{figure*}

The advantage of this representation is that it decomposes the lattice into the space group symmetry and the asu set. This allows a machine learning model to learn symmetry attributes and atomic clusters together, preserving high symmetry order in generated structures (Figure \ref{fig:asu_show}).

All four crystal representations described above satisfy translational symmetry and can be selected as the basic unit to construct the geometric graph. The entire crystal structure can then be generated through translational propagation of the lattice.

\begin{equation}
\mathbf{T}_f  = \begin{bmatrix}
1 & 0 & 0 & n_1 \\
0 & 1 & 0 & n_2 \\
0 & 0 & 1 & n_3 \\
0 & 0 & 0 & 1 \\
\end{bmatrix}
\end{equation}

\begin{equation}
\begin{bmatrix} 
u' \\ 
v' \\ 
w' \\ 
1 
\end{bmatrix}
= \mathbf{T}_f \cdot
\begin{bmatrix} 
u \\ 
v \\ 
w \\ 
1 
\end{bmatrix}
\end{equation}

where\( (u, v, w) \) and \( (u', v', w') \) are the fractional coordinates before and after translation, \( \mathbf{T}_f \) is the fractional translation matrix, and \( n_1, n_2, n_3 \in \mathbb{Z} \) are integers.

The difference between Cartesian coordinates \( \mathbf{X} \) and fractional coordinates \( \mathbf{F} = (u, v, w)  \) is straightforward. The former uses an absolute reference system, while the latter uses the crystal lattice vectors as the reference system. They can be converted between each other as follows:

\begin{align}
    \mathbf{X}&=\mathbf{LF}, \\
    \mathbf{F}&=\mathbf{L}^{-1}\mathbf{X},
\end{align}

With the atomic basis and propagation operations, any perfect crystal can be defined.

\textbf{Textural Representation}. Crystals are commonly represented using Crystallographic Information Files (CIF) \cite{hall1991crystallographic}, a standardized text-based format developed by the International Union of Crystallography (IUCr) (Figure \ref{fig:cry_rep} (II)). A CIF file encodes essential structural information, including the lattice parameters, atomic positions, symmetry operations, and space group, providing a comprehensive description of the crystal's geometry. This format facilitates the exchange and analysis of crystallographic data across various computational tools and databases. As a foundational representation, CIF files serve as the textual input format for language models, enabling the application of natural language processing techniques to crystallographic data.

\textbf{SLICES String Representation}. Methods like the Simplified Line-Input Crystal-Encoding System (SLICES) \cite{xiao2023invertible} aim to provide a sequential representation of crystal structures (Figure \ref{fig:cry_rep} (III)). Unlike traditional formats such as CIF, which rely on structured tabular data, the SLICES string linearizes key crystallographic features—such as atomic types, fractional coordinates, lattice parameters, and symmetry information—into a single sequence that satisfies both invertibility and key physical invariances. This transformation allows crystal structures to be represented in a format analogous to natural language, making them well-suited for sequence models.

\textbf{Diffraction Pattern Representation}. Powder diffraction is a physical process resulting from the interaction between coherent, monochromatic radiation and the periodic atomic arrangement of a crystal \cite{pecharsky2009fundamentals}(Figure \ref{fig:cry_rep} (IV)). Depending on the type of incident rays, powder diffraction is commonly categorized into three types: X-ray diffraction (XRD) using X-ray photons, electron diffraction, and neutron diffraction. This technique provides a rich array of structural information, encoded in the material-specific distribution of coherently scattered wave intensities, with wavelengths corresponding to interatomic lattice spacings. Although diffraction signals are inherently distributed in three-dimensional space, they are typically projected onto a one-dimensional axis to form a diffraction pattern, represented as a series of peaks. Each intensity maximum, referred to as a peak or line profile, corresponds to specific atomic arrangements along certain crystallographic directions\cite{binsimxrd,zhang2024crystallographic,hollarek2025opxrd}. As such, powder diffraction offers a physically grounded embedding of crystal structure in the form of sequential data, making it a kind of representation for machine learning. This diffraction-based sequence can be generated for any arbitrary periodic structure using toolkits such as Pysimxrd \cite{binsimxrd}.

\subsubsection{Symmetries of Crystal Structures}

\begin{figure}[t!]
\centering
\includegraphics[width=\linewidth]{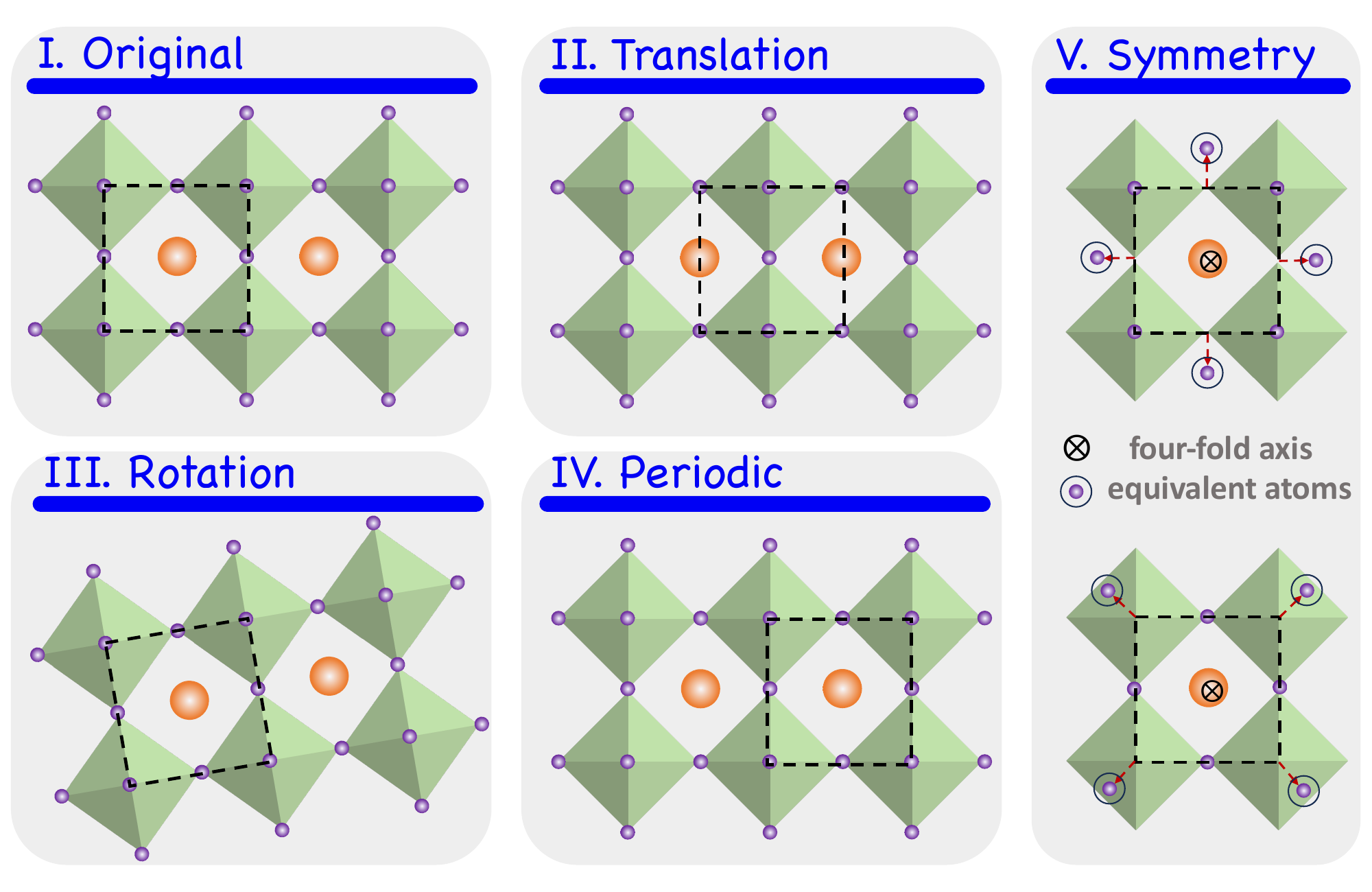}
\caption{The five types of representation invariance of crystals.}
\label{fig:equiv}
\end{figure}

The representation of a crystal structure must satisfy several fundamental invariance properties, as shown in Figure \ref{fig:equiv}:

\begin{itemize}[leftmargin=*]
    \item \textbf{Permutation invariance} ensures that the feature vector remains unchanged under any relabeling of atomic indices. This property is naturally satisfied in graph-based representations.

    \item \textbf{Translation invariance} guarantees that a uniform shift of all atomic coordinates does not affect the representation.

    \item \textbf{Rotation invariance} requires the representation to remain unchanged under any rotation of the entire structure.

    \item \textbf{Periodic invariance} ensures that the representation remains consistent under translations by any lattice vector.
\end{itemize}

These four types of invariance are essential for any representation of matter expressed via a periodic unit. In addition to these, symmetry invariance is particularly important in crystal representations.

\textbf{symmetry invariance} refers to the requirement that atoms belonging to the same point group---that is, atoms occupying the same Wyckoff position---exhibit equivalent symmetry. In such cases, the coordinates of one atom are sufficient to determine those of all other symmetrically equivalent atoms.

This invariance arises from the intrinsic symmetry of crystal structures. A symmetry operation maps an atom at coordinates $(x, y, z)$ to another atom of the same type and with identical surroundings, but located at transformed coordinates $(x', y', z')$. The common symmetry elements include:

\begin{itemize}[leftmargin=*]
    \item \textbf{Identity}: $(x, y, z) \rightarrow (x, y, z)$. This operation is always present, as it is required by group theory, the mathematical framework for symmetry.

    \item \textbf{Inversion}: $(x, y, z) \rightarrow (-x, -y, -z)$. Crystals exhibiting this symmetry are called \textit{centrosymmetric}.

    \item \textbf{Rotation axes}: A rotation axis (e.g., two-fold, three-fold) rotates the structure by an angle of $360^\circ/n$ about the axis.

    \item \textbf{Mirror planes}: These reflect atoms across a plane. For example, reflection across the $xy$-plane maps a point to its mirror image on the opposite side of the plane.
\end{itemize}

In crystallography, certain symmetry operations, such as rotation and mirroring, can be combined with translation to form special symmetry operations specific to solid materials. These combinations give rise to two additional symmetry elements: glide planes and screw axes \cite{fjellvaag1994symmetry}. While these elements are important in defining the crystal structure, they are not directly related to basis invariance in representation. A detailed discussion of these symmetry elements is beyond the scope of this section.

\begin{figure*}[h!]
\centering
\includegraphics[width=\textwidth]{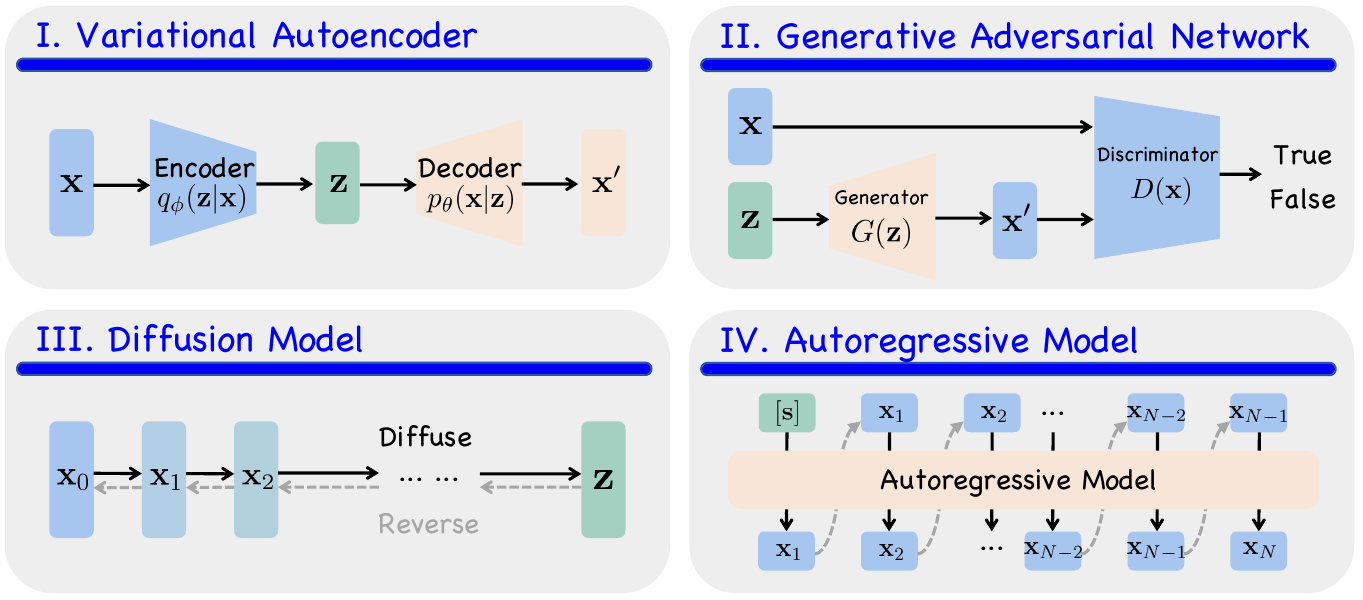}
\caption{A schematic diagram of mainstream generative model frameworks in material generation.}
\label{fig:model}
\end{figure*}

\subsection{Generative Models}

In this subsection, we will introduce the mainstream generative frameworks in modern machine learning (Figure \ref{fig:model}), including VAEs, GANs, diffusion models, and autoregressive models. Readers can skip this section and get straight to the next part if they are familiar with the background.

\subsubsection{Variational Autoencoders (VAEs)}
Variational Autoencoders (VAEs) offer a probabilistic framework for generative modeling by combining principles from variational inference with deep learning~\cite{kingma2013auto}. VAEs provide a principled approach to learning latent representations by balancing the fidelity of data reconstruction with a regularized latent space. This balance ensures that the learned manifold is smooth and amenable to generative sampling. Applications of VAEs span across image synthesis~\cite{huang2018introvae}, text-to-speech generation~\cite{kim2021conditional}, and anomaly detection~\cite{an2015variational}, among others.

A VAE consists of two main components: an encoder (or inference network) $q_\phi(\bz|\bx)$ and a decoder (or generative network) $p_\theta(\bx|\bz)$. Here, $\bx \in \mathbb{R}^{d}$ represents the observed data, and $\bz \in \mathbb{R}^{k}$ denotes the latent variables. The encoder maps data to a latent distribution, typically assumed to be Gaussian:
\begin{equation}
q_\phi(\bz|\bx) = \mathcal{N}\left(\bz; \boldsymbol{\mu}_\phi(\bx), \operatorname{diag}(\boldsymbol{\sigma}^2_\phi(\bx))\right),
\end{equation}
where $\boldsymbol{\mu}_\phi(\bx)$ and $\boldsymbol{\sigma}^2_\phi(\bx)$ are functions parameterized by $\phi$. The decoder then defines the likelihood of data given the latent variables:
\begin{equation}
p_\theta(\bx|\bz) = p(\bx; \theta, \bz),
\end{equation}
with the form of $p(\bx; \theta, \bz)$ chosen according to the nature of the data.

\noindent\textbf{Evidence Lower Bound (ELBO).}  
The training of a VAE is performed by maximizing a variational lower bound on the marginal likelihood $p_\theta(\bx)$. Formally, for each data point $\bx$, the ELBO is defined as:
\begin{equation}
\mathcal{L}(\theta, \phi; \bx) = \mathbb{E}_{\bz\sim q_\phi(\bz|\bx)}\left[\log p_\theta(\bx|\bz)\right] - D_{\text{KL}}\left(q_\phi(\bz|\bx) \,\|\, p(\bz)\right),
\end{equation}
where $p(\bz)$ is the prior distribution over the latent space, typically chosen as $\mathcal{N}(\mathbf{0}, \mathbf{I})$, and $D_{\text{KL}}(\cdot\|\cdot)$ denotes the Kullback-Leibler divergence. The first term represents the expected reconstruction loss, and the second term regularizes the encoder by measuring the divergence between the learned latent distribution and the prior.
To efficiently optimize the ELBO, the reparameterization trick~\cite{kingma2015variational} is employed to allow the gradient to back-propagate through stochastic sampling. Specifically, a latent variable $\bz$ is reparameterized as:
\begin{equation}
\bz = \boldsymbol{\mu}_\phi(\bx) + \boldsymbol{\sigma}_\phi(\bx) \odot \boldsymbol{\epsilon}, \quad \boldsymbol{\epsilon} \sim \mathcal{N}(\mathbf{0}, \mathbf{I}),
\end{equation}
where $\odot$ denotes element-wise multiplication. This formulation permits gradients to flow from the reconstruction term back to the encoder parameters $\phi$, facilitating end-to-end training.  Advanced variants, such as Conditional VAEs~\cite{kim2021conditional} and Beta-VAEs~\cite{higgins2017beta}, further refine this framework by incorporating additional conditioning variables or by explicitly controlling the trade-off between reconstruction and disentanglement.

\subsubsection{Generative Adversarial Networks}
Generative Adversarial Networks (GANs) constitute a class of generative models that formulate the generation process as a two-player minimax game between a generator and a discriminator~\cite{goodfellow2020generative}. This framework has gained significant attention due to its ability to generate high-fidelity samples in various domains such as image synthesis~\cite{zhang2018stackgan++}, image-to-image translation~\cite{isola2017image}, and audio generation~\cite{kumar2019melgan}.

Let $\bx \in \mathbb{R}^{d}$ denote data samples drawn from the real data distribution $p_{\text{data}}(\bx)$, and let $\bz \in \mathbb{R}^{k}$ be latent variables sampled from a prior distribution $p_z(\bz)$ (e.g., a multivariate Gaussian $\mathcal{N}(\mathbf{0},\mathbf{I})$). The generator $G_\theta: \mathbb{R}^{k} \to \mathbb{R}^{d}$ is parameterized by $\theta$, mapping latent codes to the data space, while the discriminator $D_\phi: \mathbb{R}^{d} \to [0,1]$, with parameters $\phi$, aims to distinguish between real and generated samples.

The training objective is cast as a minimax game:
\begin{align}
\min_{G_\theta} \max_{D_\phi} \ \mathcal{V}(D_\phi, G_\theta) &= \mathbb{E}_{\bx\sim p_{\text{data}}(\bx)}\left[\log D_\phi(\bx)\right] \\ \nonumber
&+ \mathbb{E}_{\bz\sim p_z(\bz)}\left[\log\bigl(1-D_\phi\bigl(G_\theta(\bz)\bigr)\bigr)\right].
\end{align}
In this formulation, the discriminator attempts to maximize the probability of correctly classifying both real and generated samples, whereas the generator strives to minimize $\log\bigl(1-D_\phi(G_\theta(\bz))\bigr)$ such that the synthesized samples become indistinguishable from real data. Under ideal conditions, the optimal discriminator is given by:
\begin{equation}
D^*(\bx) = \frac{p_{\text{data}}(\bx)}{p_{\text{data}}(\bx) + p_{G}(\bx)},
\end{equation}
where $p_{G}$ denotes the distribution of the generated samples induced by $G_\theta$. At equilibrium, when $p_{G} = p_{\text{data}}$, the discriminator cannot distinguish between real and generated data, and the training objective is minimized. 

Due to the adversarial nature of GAN training, the balance between $G_\theta$ and $D_\phi$ is crucial for convergence. Numerous variants have been introduced to stabilize training, such as Wasserstein GAN~\cite{arjovsky2017wasserstein}, which replaces the original loss with a Wasserstein distance-based objective:
\begin{equation}
\min_{G_\theta} \max_{D_\phi \in \mathcal{D}} \ \mathbb{E}_{\bx\sim p_{\text{data}}(\bx)}\left[D_\phi(\bx)\right] - \mathbb{E}_{\bz\sim p_z(\bz)}\left[D_\phi\bigl(G_\theta(\bz)\bigr)\right],
\end{equation}
where $\mathcal{D}$ is the set of 1-Lipschitz functions enforced by weight clipping or gradient penalty~\cite{gulrajani2017improved}. Such modifications aim to improve gradient propagation and mitigate mode collapse. Despite their conceptual elegance, GANs require careful hyperparameter tuning and architectural design to ensure convergence in practical applications.

\subsubsection{Diffusion Models}
Owing to their robust theoretical basis and notable empirical effectiveness, diffusion models have gained prominence as a powerful class of generative models. They are especially advantageous for material and molecular generation tasks, where data typically exhibit high dimensionality and intricate structure~\cite{wang2025survey}.

\noindent\textbf{Denoising Diffusion Probabilistic Models (DDPMs).}
DDPM~\cite{Diffusion,DDPM} represent a foundational approach in the diffusion model family, utilizing a fixed noise schedule to iteratively denoise data. The framework is built on two Markov chains: one for the forward (noising) process and another for the reverse (denoising) process.

Beginning with clean data $\bx_0$, the forward process progressively corrupts it into a sequence $\bx_1, \bx_2, \ldots, \bx_T$ through a series of Gaussian transitions defined as:
\begin{equation}
q(\bx_t|\bx_{t-1}) = \mathcal{N}\left(\bx_t; \sqrt{\alpha_t}\bx_{t-1}, (1-\alpha_t)\bI\right),
\end{equation}
where each $\alpha_t \in (0, 1)$ controls the noise level at step $t$. Here, $\mathcal{N}(\bx; \boldsymbol{\mu}, \mathbf{\Sigma})$ denotes a Gaussian distribution with mean $\boldsymbol{\mu}$ and covariance $\mathbf{\Sigma}$.
A key property of this process is that the noisy data at any time step $t$ can be expressed directly in terms of the original input $\bx_0$:
\begin{equation}
q(\bx_t|\bx_0) = \mathcal{N}\left(\bx_t; \sqrt{\bar{\alpha}_t}\bx_0, (1-\bar{\alpha}_t)\bI\right),
\label{eq:forward-property}
\end{equation}
where $\bar{\alpha}_t := \prod_{i=1}^t \alpha_i$. This yields a closed-form expression: $\bx_t = \sqrt{\bar{\alpha}_t}\bx_0 + \sqrt{1-\bar{\alpha}_t}\boldsymbol{\epsilon}$, with $\boldsymbol{\epsilon} \sim \mathcal{N}(\mathbf{0}, \bI)$. Typically, the noise schedule is chosen so that $\bar{\alpha}_T \approx 0$, ensuring that the final state $\bx_T$ approximates standard Gaussian noise, which serves as the starting point for the reverse trajectory.

The reverse transition is modeled using neural networks that parameterize the mean and covariance functions, denoted by $\boldsymbol{\mu}_{\theta}$ and $\mathbf{\Sigma}_{\theta}$, respectively:
\begin{equation}
p_\theta(\bx_{t-1}|\bx_t) = \mathcal{N}\left(\bx_{t-1}; \boldsymbol{\mu}_{\theta}(\bx_t, t), \mathbf{\Sigma}_{\theta}(\bx_t, t)\right),
\label{eq:reverse}
\end{equation}
where $\theta$ represents the set of learnable parameters.
Training involves maximizing the likelihood of the observed data $\bx_0$, which is equivalent to minimizing the evidence lower bound of the negative log-likelihood $\mathbb{E}[-\log p_\theta(\bx_0)]$.

DDPM simplifies the covariance term $\mathbf{\Sigma}_\theta$ in \cref{eq:reverse} by fixing it to a time-dependent diagonal matrix, specifically $\tilde{\beta}_t \bI$, where $\tilde{\beta}_t = \frac{1 - \bar{\alpha}_{t-1}}{1 - \bar{\alpha}_t}(1 - \alpha_t)$ controls the noise level at each timestep. Furthermore, the mean $\boldsymbol{\mu}_\theta$ is reformulated using a learnable noise prediction network $\boldsymbol{\epsilon}_\theta$ as follows:
\begin{equation}
\boldsymbol{\mu}_\theta(\bx_t, t) = \frac{1}{\sqrt{\alpha_t}} \left(\bx_t - \frac{1 - \alpha_t}{\sqrt{1 - \bar{\alpha}_t}} \boldsymbol{\epsilon}_\theta(\bx_t, t) \right),
\end{equation}
where $\boldsymbol{\epsilon}_\theta(\bx_t, t)$ estimates the noise $\boldsymbol{\epsilon}$ added to the clean data $\bx_0$ at step $t$.
Leveraging the property in \cref{eq:forward-property} and omitting constant weighting factors, the training objective reduces to a simplified denoising score matching loss:
\begin{equation}
\mathbb{E}_{t,\bx_0,\boldsymbol{\epsilon}} \left[\left\| \boldsymbol{\epsilon} - \boldsymbol{\epsilon}_\theta \left( \sqrt{\bar{\alpha}_t} \bx_0 + \sqrt{1 - \bar{\alpha}_t} \boldsymbol{\epsilon}, t \right) \right\|^2 \right].
\end{equation}
During inference, sampling starts from pure Gaussian noise $\bx_T \sim \mathcal{N}(\mathbf{0}, \bI)$, and the reverse process iteratively denoises it. For $t = T, T-1, \ldots, 1$, the update rule is:
\begin{equation}
\bx_{t-1} \leftarrow \frac{1}{\sqrt{\alpha_t}} \left( \bx_t - \frac{1 - \alpha_t}{\sqrt{1 - \bar{\alpha}_t}} \boldsymbol{\epsilon}_\theta(\bx_t, t) \right) + \sigma_t \bz,
\end{equation}
where $\bz \sim \mathcal{N}(\mathbf{0}, \bI)$ for $t = T, \ldots, 2$, and $\bz = \mathbf{0}$ when $t = 1$.

DDPM has found extensive use in material generation, particularly for tasks involving continuous data representations. Due to its formulation in continuous space, it is especially well-suited for generating continuous 3D material structures, as demonstrated in works such as DiffCSP~\cite{jiao2023crystal}, UniMat~\cite{yang2023scalable}, SymmCD~\cite{levy2025symmcd}, and MatterGen~\cite{zeni2025generative}.

Score Matching with Langevin Dynamics (SMLDs)~\cite{SMLD} were developed in parallel with DDPMs. These models leverage score-based methods~\cite{ScoreMatching} to estimate the gradient of the data distribution—known as the Stein score—and integrate this with Langevin dynamics for sampling. Through denoising score matching~\cite{DenoisingScoreMatching}, a neural network is trained to approximate the Stein score across varying noise levels, enabling iterative generation of samples.

SMLDs have also been applied to material generation, with examples including CDVAE~\cite{xie2021crystal} and MOFDiff~\cite{fu2023mofdiff}. Importantly, SMLDs and DDPMs are theoretically equivalent, as both can be interpreted as discrete approximations of underlying stochastic differential equations (SDEs).

SDEs~\cite{SDE} generalize the discrete formulations of DDPMs and SMLDs to continuous time, providing enhanced flexibility and control over the generative process. In this framework, the forward trajectory is described by a SDE, while the reverse process is represented either as its time-reversed counterpart or an equivalent probability flow ordinary differential equation (ODE). Notably, specific SDE variants, such as variance-preserving (VP) and variance-exploding (VE) SDEs, correspond to DDPMs and SMLDs, respectively.

\noindent\textbf{Flow Matching (FM).}
Flow Matching is a recently proposed generative modeling framework that directly learns a continuous-time transformation between a simple base distribution (e.g., Gaussian noise) and the target data distribution. Unlike DDPM, SMLD, or SDE-based methods, which rely on stochastic processes or discrete approximations, FM focuses on learning a deterministic flow field that maps samples from the base distribution to the data distribution in a single continuous trajectory. This approach is inspired by the theory of optimal transport and leverages the concept of matching probability flows.

The core idea of FM is to parameterize the velocity field of the transformation using a neural network. Let $ \bx(t) $ represent the state of a sample at time $t \in [0, 1]$, where $t=0$ corresponds to the target data distribution and $t=1$ corresponds to the base distribution. The evolution of $ \bx(t) $ is governed by an ordinary differential equation (ODE):
\begin{equation}
\frac{d\bx}{dt} = \bv_\theta(\bx, t),
\end{equation}
where $ \bv_\theta(\bx, t) $ is the velocity field parameterized by a neural network with learnable parameters $ \theta $. The goal is to train $ \bv_\theta $ such that the flow induced by the ODE transforms the base distribution into the target distribution.

To train the model, FM minimizes a loss function that measures the discrepancy between the learned velocity field $ \bv_\theta(\bx, t) $ and the ideal velocity field that would perfectly match the probability flows. This is achieved by solving a least-squares problem based on the continuity equation, which describes the conservation of probability mass during the transformation. Specifically, the objective function is:
\begin{equation}
\mathbb{E}_{t, \bx_t, \bx_0} \left[ \left\| \bv_\theta(\bx_t, t) - \mathbf{v}^*(\bx_t, t) \right\|^2 \right],
\end{equation}
where $ \mathbf{v}^*(\bx_t, t) $ is the ideal velocity field derived from the target distribution and the forward process.

In practice, training samples are typically generated by linearly interpolating between data samples $\bx_0 \sim p_{\text{data}}$ and base samples $\bx_1 \sim p_{\text{base}}$, where $p_{\text{base}}$ is typically a standard Gaussian. For each pair $(\bx_0, \bx_1)$, a time $t \in [0, 1]$ is sampled uniformly, and the interpolated point is computed as:
\begin{equation}
\bx_t = (1 - t) \bx_0 + t \bx_1.
\end{equation}
Then the ideal velocity can be computed as $\bv^*(\bx_t, t) = \bx_1 - \bx_0$.
Once the velocity field $ \bv_\theta(\bx, t) $ is trained, samples can be generated by solving the learned ODE in reverse, starting from the base distribution. This process is deterministic and does not require iterative denoising or stochastic sampling, making FM computationally efficient compared to diffusion-based methods.

FM has shown promise in material generation~\cite{miller2024flowmm,kim2024mofflow,luo2024crystalflow}, due to its ability to model complex distributions with high efficiency and flexibility. By directly learning the flow field, FM avoids the need for discretizing stochastic processes, offering a more straightforward and interpretable approach to generative modeling.

\begin{figure*}[h!]
\centering
\includegraphics[width=\textwidth]{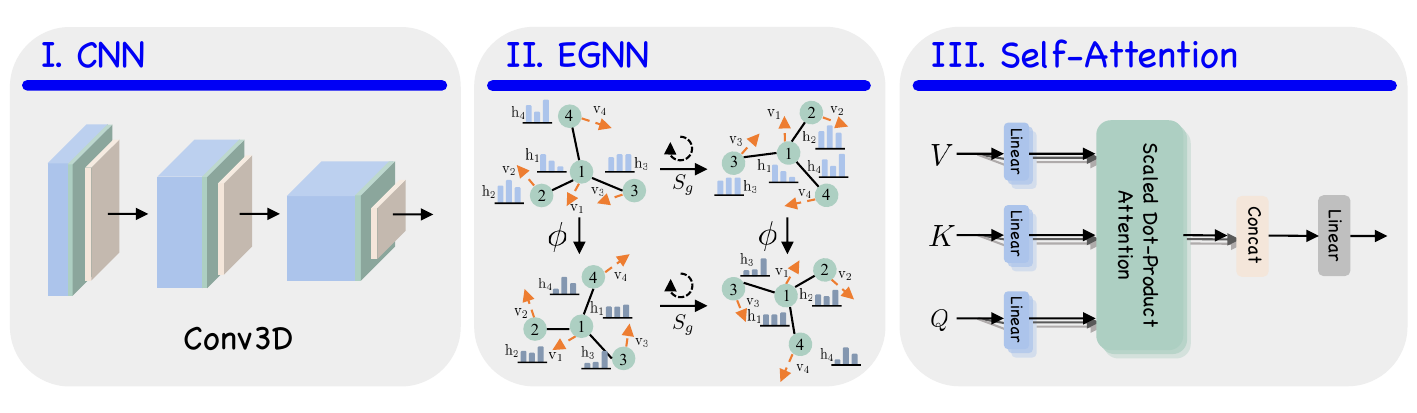}
\caption{A general illustration of model backbones: CNN, GNN, and Transformers.}
\label{fig:encoder}
\end{figure*}

\subsubsection{Autoregressive Models}

Autoregressive modeling has been a huge success in the field of natural language processing (NLP). Recently, autoregressive models have emerged as a significant area of focus in computer vision \cite{lee2022autoregressive}, protein design \cite{shin2021protein}, and etc. These models operate by factorizing the joint probability distribution of a sequence into a product of conditional probabilities, allowing each token or data point to be predicted based on the preceding ones. Formally, given a sequence:
\begin{equation}
    \mathbf{x} = x_1,x_2,\ldots,x_T,
\end{equation}
the autoregressive approach models the probability as:
\begin{equation}
    p(\mathbf{x})=\prod_{t=1}^Tp(x_i\mid x_1,x_2\ldots,x_{t-1}; \theta),
\end{equation}
where $p(x_t\mid x_1,x_2,\ldots, x_{t-1};\theta)$ represents the probability of the current elements $x_t$ conditioned on all previous elements in the sequence, with $\theta$ denoting the model parameters. This property makes them highly effective for tasks such as language modeling, where predicting the next word in a sentence requires contextual understanding of the preceding words. The training objective is to minimize the negative log-likelihood (NLL) loss, which is formulated as:
\begin{equation}
    \mathcal{L}(\theta)=-\sum_{t=1}^T\log p(x_t\mid x_1, x_2,\ldots, x_{t-1};\theta).
\end{equation}
By minimizing this objective, the model learns to predict the next token, effectively capturing the underlying structure of the sequence data.

\section{Model Backbone}
\label{sec:backbone}

The model backbone is critical in determining the ability to learn and generate latent material representations (Figure \ref{fig:encoder}). This section introduces the primary neural network architectures, Convolutional Neural Networks (CNNs)~\cite{o2015introduction}, Graph Neural Networks (GNNs)~\cite{DBLP:conf/icml/GilmerSRVD17,kipf2016semi,gasteiger2020directional}, and Transformers~\cite{vaswani2017attention}, with the evolution of material input types, from grid-based tensors to geometric graphs, and recent text-like sequences.

\subsection{Convolutional Neural Networks}
Early material generation relied on grid-structured representations, such as voxelized grids~\cite{noh_inverse_2019,long_constrained_2021} or unit cell parameters~\cite{kim_generative_2020}, which naturally align with tensors. For a 3D tensor input $\mathbf{H}$, CNNs became the primary backbone due to their ability to capture local spatial patterns through hierarchical convolutions:
\begin{equation}
    \mathbf{H}^{(l+1)} = \sigma(\mathbf{W}^{(l)}*\mathbf{H}^{(l)} + \mathbf{b}^{(l)}),
\end{equation}
where $*$ denotes convolution, $\mathbf{W}^{(l)}$ and $\mathbf{b}^{(l)}$ are filters and biases. $\sigma$ is an activation function.
However, CNNs require rigid grid inputs, forcing irregular atomic arrangements into tensor formats. This preprocessing loses critical geometric details (e.g., bond angles) and struggles with physical symmetry.

\subsection{Graph Neural Networks}
GNNs~\cite{gasteiger2021gemnet,gasteiger2020directional} have emerged as the dominant backbone for material generation due to their natural compatibility with crystal structures represented as geometric graphs. By treating atoms as nodes and interatomic interactions as edges, GNNs inherently respect permutation invariance and local atomic environments~\cite{keriven2019universal,garg2020generalization}. Equivariant GNNs~\cite{han2024survey} are further proposed to respect the inherent symmetries of crystal structures, including permutation invariance, rotational/translational invariance, and periodicity. Since the composition of equivariant functions is again equivariant~\cite{zheng2024relaxing,liu2023segno}, equivariance is achieved via stacking multiple equivariant message-passing layers. Thus, their output will change in the same way as the input changes. Formally, they consist of two steps:
\begin{equation}
\begin{aligned}
    \mathbf{M}_{ij}^{(l)}  &= \mu(\mathbf{H}_{i}^{(l)}, \mathbf{H}_{j}^{(l)}, e_{ij}), \\
    \mathbf{H}_{i}^{(l+1)} &= \nu(\mathbf{H}_{i}^{(l)}, \sum_{j\in \mathcal{N}_i}\mathbf{M}_{ij}^{(l)}),
\end{aligned}
\end{equation}
where $\mathbf{M}_{ij}^{(l)}$ denotes the $l$-th layer message embedding between node $i$ and node $j$. $\mathcal{N}_i$ collects the neighbors of node $i$. $e_{ij}$ is the edge features. $\mu$ and $\nu$ are the equivariant message embedding function and node state updating function, respectively. For $SO(3)$ equivariance, common backbones are Gemnet~\cite{gasteiger2021gemnet}, Dimenet~\cite{gasteiger2020directional}, and EGNN~\cite{satorras2021n}. Periodicity equivariance ensures that generated crystal structures inherently respect the symmetry and infinite repetition of atomic arrangements in periodic lattices. Such property can be ensured by employing fractional coordinates~\cite{jiao2023crystal}, encoding relative positions between atoms using Fourier transforms~\cite{jiao2024space}, and modeling atomic interactions across periodic boundaries~\cite{xie2021crystal}.

\subsection{Transformer}

Transformer architectures~\cite{vaswani2017attention} have revolutionized sequential data modeling across domains, and their application~\cite{touvron2023llama, achiam2023gpt, liu2024deepseek} in materials generation has grown rapidly. Specifically, for an input sequence (e.g., graphs or texts) represented as a matrix $\mathbf{X}\in\mathbb{R}^{T\times d_{\text{model}}}$ (where $T$ is the sequence length and $d_\text{model}$ is the model's dimensionality), the input embedding are projected into $h$ independent subspaces, where $h$ is the number of attention heads. This is achieved by using separate learned weight matrices for each $i$-th head:
\begin{align}
    \mathbf{Q}_i=\mathbf{XW_Q}^{(i)}, \mathbf{K}_i=\mathbf{XW_K}^{(i)}, \mathbf{V}_i=\mathbf{XW_V}^{(i)},
\end{align}
where $i=1,2,\ldots,h$, $\mathbf{Q}_i,\mathbf{K}_i,\mathbf{V}_i$ car query, key, and value matrices respectively, $\mathbf{W_Q}^{(i)}$, $\mathbf{W_K}^{(i)}$, and $\mathbf{W_V}^{(i)}\in\mathbb{R}^{d_\text{model}\times d_k}$ are the learned weight matrices for the $i$-the attention head, and $d_k$ is typically set to $d_\text{model}/h$ so that the total computation cost remains constant. The attention weights are calculated as scaled dot-products between the queries and keys, followed by a softmax operation to normalize them:
\begin{equation}
    \mathbf{H}_i=\text{softmax}\Bigg(\frac{\mathbf{Q}_i\mathbf{K}_i^T}{\sqrt{d_k}}\Bigg)\mathbf{V}_i,
\end{equation}
where $\mathbf{H}_i\in\mathbb{R}^{T\times d_k}$ is the output of the $i$-th attention head. Then, the final output is calculated by concatenating the outputs for $h$ heads along the features dimension:
\begin{equation}
    \mathbf{O}=\text{Concat}(\mathbf{H}_1,\mathbf{H}_2,\ldots,\mathbf{H}_h)\cdot\mathbf{W_O},
\end{equation}
where $\mathbf{W_O}\in\mathbb{R}^{(h\cdot d_k)\times d_\text{model}}$ is a linear transformation.The resulting matrix $\mathbf{O}\in\mathbb{R}^{T\times d_\text{model}}$ represents the final output of the multi-head attention mechanism, which is then passed to subsequent layers in the transformer architecture.

Current methods leverage Transformers for two paradigms. The first approaches~\cite{flam2023language,flam2023language,joshi2025all,tangsongcharoen2025crystalgrw} encode material representations such as geometric graphs or sequentialized data, exemplified by models like xyztransformer~\cite{flam2023language} and Mat2Seq~\cite{yan2024invariant}, which serialize crystallographic information (e.g., CIF files) into token sequences for autoregressive generation of atomic coordinates and lattice parameters. The second methods~\cite{yang2024generative,gruver2024fine,xia2025naturelm,ding2024matexpert,gan2025large,sriram2024flowllm}, on the other hand, leverage natural language to drive discovery: GenMS~\cite{yang2024generative} translates textual prompts into chemical formulas for diffusion-based structure generation, while CrystalLLM~\cite{gruver2024fine} fine-tunes LLaMA-2 to produce symmetry-compliant crystals from compositional instructions. Advanced multimodal systems like NatureLM~\cite{xia2025naturelm} unify scientific text, formulas, and crystallographic data, enabling cross-domain tasks such as generating hypothetical superconductors from theoretical hypotheses.

\section{Taxonomy}
\label{sec:taxonomy}

\begin{figure*}[h!]
\centering
\includegraphics[width=\textwidth]{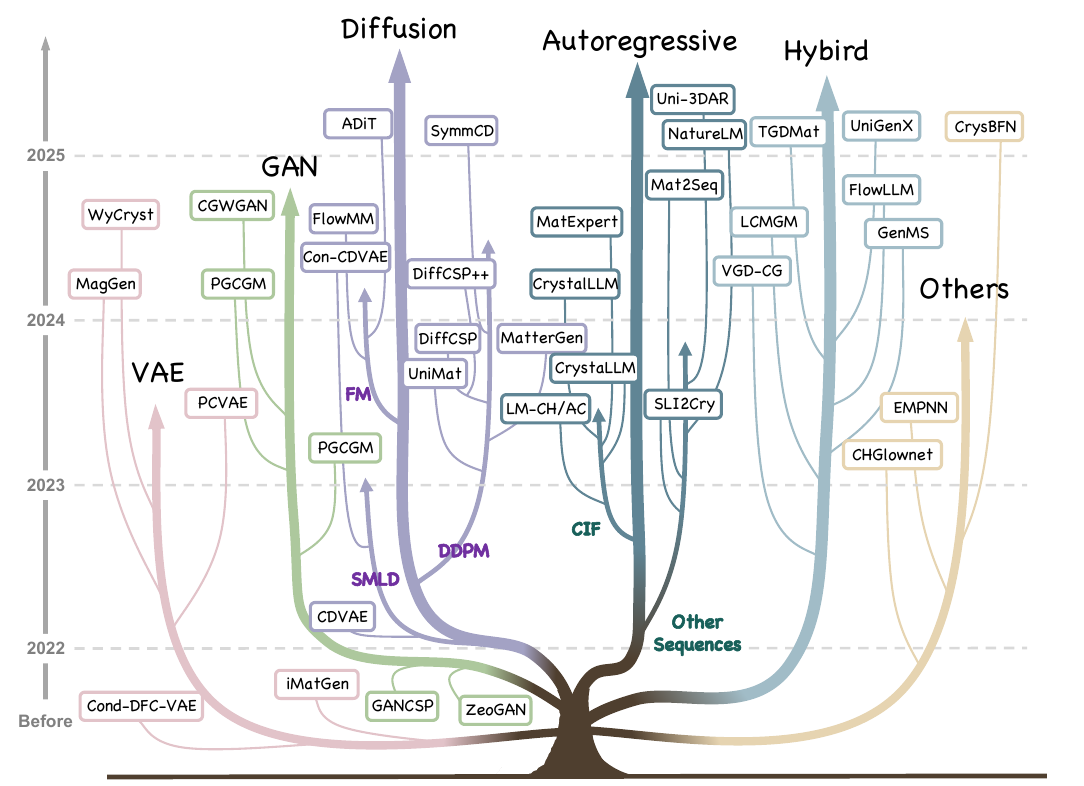}
\caption{A timeline overview of material generative models developed by different methods.}
\label{fig:timeline}
\end{figure*}

This section presents a comprehensive introduction of cutting-edge approaches to materials generation. We categorize existing methods into six major type, including VAE-based, GAN-based, Diffusion-based, Autoregressive-based, Hybird models, and others (Table \ref{tab:taxonomy}). Furthermore, to provide readers with a clearer understanding of the evolution of this research area, we have constructed a timeline highlighting representative methods (Figure \ref{fig:timeline}).

\begin{table*}[t]
\caption{\textbf{A comparison of deep learning-based material generation methods with respect to the provided taxonomy.} \textbf{Method}: The specific methods used in the material generative models. \textbf{Materials}: The target objects generated by the model include various types of material, such as inorganic crystals, superconductors, metal-organic frameworks, and etc. \textbf{Backbone}: The architecture used by the model to produce intermediate representations. \textbf{Condition}: The conditions can be used in conditional generation. \textbf{Size}: The parameter size of the generative models.}
\resizebox{\linewidth}{!}{
\begin{tabular}{l|l|llllll}
\toprule
& \textbf{Model} & \textbf{Method} & \textbf{Materials} & \textbf{Backbone} & \textbf{Condition} & \textbf{Size} & \textbf{Code \& Year} \\
\midrule

\multirow{6}{*}{\rotatebox{90}{VAE}} 

& iMatGen \cite{noh_inverse_2019}          & VAE & Inorganic Crystals & CNN & Composition, Property & $\sim$7M & \href{https://github.com/kaist-amsg/imatgen}{2019}\\

& Cond-DFC-VAE \cite{court_3-d_2020}       & VAE & Inorganic Crystals&  CNN & Property & - & \href{https://github.com/by256/icsg3d}{2020}\\

& FTCP \cite{ren_invertible_2022}          & VAE & Inorganic Crystals &  CNN& Property & -  & \href{https://github.com/PV-Lab/FTCP}{2022}\\
& PCVAE \cite{liu2023pcvae}                & VAE & Inorganic Crystals &  MLP & Composition & $\sim$3M & \href{https://github.com/zjuKeLiu/PCVAE}{2023}\\
& WyCryst \cite{zhu_wycryst_2024}          & VAE & Inorganic Crystals&  CNN & Composition, Property   & -  & \href{https://github.com/RaymondZhurm/WyCryst}{2024}\\
& MagGen \cite{mal_maggen_2024}            & VAE &Permanent Magnets&  - & Property & -& 2024\\
\midrule

\multirow{10}{*}{\rotatebox{90}{GAN}} 
& GANCSP \cite{kim_generative_2020}        & GAN & Inorganic Crystals &  CNN & Composition & $\sim$4M & \href{https://github.com/kaist-amsg/Composition-Conditioned-Crystal-GAN}{2020}\\

& CCDCGAN \cite{long_constrained_2021}     & GAN & Inorganic Crystals &  CNN & Composition, Property& -& 2021\\

& ZeoGAN \cite{kim_inverse_2020}           & GAN & Zeolites &  CNN & Property& $\sim$39M& \href{https://github.com/good4488/ZeoGAN}{2020}\\
& PGCGM \cite{zhao_physics_2023}           & GAN & Inorganic Crystals &  CNN & Composition, Space Group& $\sim$5.5M & \href{https://github.com/MilesZhao/PGCGM}{2023}\\
& GAN-DDLSF \cite{chen_crystal_2024}       & GAN & Gallium Nitride& - & Composition & - & 2024\\

& NSGAN \cite{li_nsgan_2024}               & GAN & Aluminium Alloys& MLP & Composition, Property& $\sim$5K& \href{https://github.com/anucecszl/NSGAN_aluminium}{2024}\\
& MatGAN \cite{dan2020generative}          & GAN & Inorganic Crystals & CNN & Property & - & 2020 \\
& CubicGAN \cite{zhao2021high}             & GAN & Cubic Crystal       & CNN & Composition, Space Group & - & 2021 \\

& DeepCSP \cite{ye2024organic}             & GAN & Organic Crystal     &  GCN & Composition & - & 2024 \\
& CGWGAN \cite{su2024cgwgan}               & GAN & Inorganic Crystals &  MLP & Composition & 0.38M & \href{https://github.com/WPEM/CGWGAN}{2024} \\
\midrule

\multirow{21}{*}{\rotatebox{90}{Diffusion}} 
& CDVAE \cite{xie2021crystal}              & SMLD & Inorganic Crystals & DimeNet+GemNet & Property & 4.5M & \href{https://github.com/txie-93/cdvae}{2021} \\
& Cond-CDVAE \cite{luo_deep_2024}          & SMLD & Inorganic Crystals & DimeNet+GemNet & Composition, Property& 4M/86M& \href{https://github.com/ixsluo/cond-cdvae}{2024}\\ 
& Con-CDVAE \cite{ye2024cdvae}             & SMLD & Inorganic Crystals & DimeNet+GemNet & Composition, Property & $\sim$5M & \href{https://github.com/cyye001/Con-CDVAE}{2024}\\

& P-CDVAE \cite{kaszuba_compositional_2023} & SMLD& Inorganic Crystals& DimeNet+GemNet  & Composition, Property & - & 2023\\
& LCOMs \cite{qi_latent_2023}              & SMLD & Inorganic Crystals& DimeNet+GemNet & Composition & 4.5M & 2023\\
& StructRepDiff \cite{sinha_representation-space_2024} & DDPM & Inorganic Crystals & U-Net  & - & 1$\sim$10M & 2024\\
& DiffCSP \cite{jiao2023crystal}           & DDPM & Inorganic Crystals & Periodic GNN & Composition & 12.3M & \href{https://github.com/jiaor17/DiffCSP}{2023} \\
& UniMat \cite{yang2023scalable}           & DDPM & Inorganic Crystals & U-Net & Composition, Property & - & \href{https://unified-Crystals.github.io/unimat/}{2023} \\
& DiffCSP++ \cite{jiao2024space}           & DDPM & Inorganic Crystals & Periodic GNN & Composition, Space Group &  12.3M & \href{https://github.com/jiaor17/DiffCSP-PP}{2024} \\
& GemsDiff \cite{klipfel2024vector}        & DDPM & Inorganic Crystals & GemsNet & Composition & 2.8M & \href{https://github.com/aklipf/gemsdiff}{2024} \\
& EquiCSP \cite{lin2024equivariant}        & DDPM & Inorganic Crystals & Periodic GNN & Composition & 12.3M & \href{https://github.com/EmperorJia/EquiCSP}{2024} \\
& FlowMM \cite{miller2024flowmm}           & RFM  & Inorganic Crystals & Periodic GNN & Composition & 12.3M & \href{https://github.com/facebookresearch/flowmm}{2024} \\
& SuperDiff \cite{yuan2024diffusion}       & DDPM & Superconductors     & U-Net & Composition, Property & - & \href{https://github.com/sdkyuanpanda/SuperDiff}{2024} \\
& SymmCD \cite{levy2025symmcd}             & DDPM & Inorganic Crystals & Periodic GNN & Composition, Space Group & 12.3M & \href{https://github.com/sibasmarak/SymmCD}{2025} \\
& MOFDiff \cite{fu2023mofdiff}             & SMLD & Metal-organic Frameworks & GemNet & Property & 27.2M & \href{https://github.com/microsoft/MOFDiff}{2023} \\
& MatterGen \cite{zeni2025generative}      & DDPM & Inorganic Crystals & GemNet & Composition, Space Group, Property & 46.8M & \href{https://github.com/microsoft/mattergen}{2025} \\
& MOFFlow \cite{kim2024mofflow}            & RFM  & Metal-organic Frameworks & EGNN+OpenFold & Composition & 22.5M & \href{https://github.com/nayoung10/MOFFlow}{2024} \\
& CrystalFlow \cite{luo2024crystalflow}    & RFM  & Inorganic Crystals & Periodic GNN & Composition, Property & 12.3M & \href{https://github.com/ixsluo/CrystalFlow}{2024} \\
& ADiT \cite{joshi2025all}                 & LFM  & Atomic Systems      & Transformer & - & 32M/130M/450M & \href{https://github.com/facebookresearch/all-atom-diffusion-transformer}{2025} \\
& DAO \cite{wu2025siamese}                 & DDPM & Inorganic Crystals, Superconductors & Periodic Transformer & Composition, Property & - & 2025 \\
& CrystalGRW \cite{tangsongcharoen2025crystalgrw} & GRW & Inorganic Crystals & EquiformerV2 & Property & 34.9M & \href{https://github.com/trachote/crystalgrw}{2025} \\
\midrule

\multirow{10}{*}{\rotatebox{90}{Autoregressive}} 
& G-SchNet \cite{gebauer2019symmetry}      & NTP  & Atomic Systems      & CNN   & Composition & - & \href{https://github.com/atomistic-machine-learning/G-SchNet}{2019} \\
& xyztransformer \cite{flam2023language}         & NTP  & Atomic Systems      & Transformer & Composition & 1$\sim$100M & \href{https://github.com/danielflamshep/xyztransformer}{2023} \\
& CrystaLLM \cite{antunes2024crystal}      & NTP  & Inorganic Crystals & GPT-2 & Composition & 25M & \href{https://github.com/lantunes/CrystaLLM}{2024} \\
& CrystalLLM \cite{gruver2024fine}         & NTP  & Inorganic Crystals & LLaMA-2 & Composition, Text & 7B/13B/70B & \href{https://github.com/facebookresearch/crystal-text-llm}{2024} \\
& SLI2Cry \cite{xiao2023invertible}        & NTP & Inorganic Crystals & GRU & Composition, Property & - & \href{https://github.com/xiaohang007/SLICES}{2023} \\
& Mat2Seq \cite{yan2024invariant}          & NTP  & Inorganic Crystals & GPT-2 & Composition & 25M/200M & \href{https://github.com/YKQ98/Mat2Seq}{2024} \\
& MatExpert \cite{ding2024matexpert}       & RAG  & Inorganic Crystals & LLaMA-2/3 & Composition, Text & 8B/70B & \href{https://github.com/BangLab-UdeM-Mila/MatExpert}{2024} \\
& NatureLM \cite{xia2025naturelm}          & NTP & Atomic Systems & Transformer & Composition, Space Group, Property & 1B/8B/46.7B & \href{https://naturelm.github.io/}{2025} \\
& MatLLMSearch \cite{gan2025large}         & NTP & Inorganic Crystals & LLaMA-3.1 & Composition & 70B & \href{https://github.com/JingruG/MatLLMSearch}{2025} \\
& Uni-3DAR \cite{lu2025uni}                & MNTP & Atomic Systems & Transformer & Composition, Text & 90M & \href{https://github.com/dptech-corp/Uni-3DAR}{2025} \\
\midrule

\multirow{7}{*}{\rotatebox{90}{Hybird}} 
& FlowLLM \cite{sriram2024flowllm}         & LLM+RFM & Inorganic Crystals & LLaMA-2+GNN & Composition & 70B & \href{https://github.com/facebookresearch/flowmm}{2024}\\
& GenMS \cite{yang2024generative}          & LLM+DM & Inorganic Crystals & Gemini+Transformer& Composition & - & 2024 \\
& TGDMat \cite{das2025periodic}            & LM+DM & Inorganic Crystals & SciBERT+EGNN & Composition, Space Group & - & \href{https://github.com/kdmsit/TGDMat}{2025} \\
& UniGenX \cite{zhang2025unigenx}          & NTP+DDPM & Atomic Systems & Transformer+MLP  & Composition & 100M/400M & 2025 \\
& LCMGM \cite{chenebuah_deep_2024}         & VAE+GAN & Inorganic Crystals & CNN & Crystal System & - & \href{https://github.com/chenebuah/LCMGM}{2024} \\
& VGD-CG \cite{qin_inverse_2024}           & VAE+GAN+DDPM & Inorganic Crystals, Semiconductor & U-Net & Composition & - & \href{https://github.com/stupidcloud/VGD-CG}{2024} \\
& DP-CDVAE \cite{pakornchote_diffusion_2024}  & DDPM+VAE & Crystal Structures& GemNet & Composition, Space Group, Property & -& \href{https://github.com/trachote/dp-cdvae}{2024}\\ 
\midrule

\multirow{3}{*}{\rotatebox{90}{Others}} 
& EMPNN \cite{klipfel2023equivariant}      & MPNN & Inorganic Crystals & MPNN  & Composition, Noisy Structure & - & \href{https://github.com/aklipf/pegnn}{2023} \\
& CrysBFN \cite{wu2025periodic}            & BFN  & Inorganic Crystals & Periodic GNN  & Composition & 12.3M & \href{https://github.com/wu-han-lin/CrysBFN}{2025} \\
& CHGlownet \cite{nguyen2023hierarchical} & GFlowNet  & Inorganic Crystals &  GCN+MLP& Composition, Space Group, Property & - & 2023 \\
\bottomrule
\end{tabular}}
\label{tab:taxonomy}
\end{table*}

\subsection{VAE-based Models}
VAEs have offered a valuable pathway for materials discovery, primarily due to their capacity to learn a continuous and often structured latent space representation of crystal data. This continuous embedding facilitates the sampling of novel candidates and interpolation between known structures, making VAEs a flexible framework for inverse design. Early works leveraging VAEs for crystal generation often adopted grid-based representations. For example, iMatGen \cite{noh_inverse_2019} introduced a hierarchical, two-step VAE architecture for the inverse design of vanadium oxides (V$_x$O$_y$), which compressed image-based fingerprints of the unit cell and atomic basis before learning a VAE over these representations. Building upon the concept of grid-based inputs, Court et al. \cite{court_3-d_2020} subsequently propose Cond-DFC-VAE operating directly on 3D voxelized electron-density maps. Conditioned on formation energy, their model extended applicability to diverse material classes like binary alloys, ternary perovskites, and Heusler compounds. Key innovations in their pipeline included the integration of coordinate convolution for accurate unit-cell dimension recovery, a U-Net for segmenting generated electron-density maps into discrete atomic structures, and a Crystal Graph Convolutional Neural Network \cite{xie2018crystal} to concurrently predict multiple physical properties of the de novo generated crystals, thus offering a more holistic generative and predictive capability.

To achieve more general inverse design capabilities, Ren et al. \cite{ren_invertible_2022} develop the FTCP framework, which employs a VAE with a novel crystallographic representation that combines real-space (CIF-like) features with reciprocal-space features derived from Fourier-transformed elemental properties. This richer representation, coupled with a target-learning branch in the VAE, enabled conditioning on multiple target properties simultaneously, such as formation energy, bandgap, thermoelectric power factor. This framework achieves successful generation of diverse, novel crystals validated by DFT calculations. To address the challenge of multiple stable crystal structures existing for a single chemical formula, Liu et al. \cite{liu2023pcvae} introduces the PCVAE, which encodes the various possible crystal structures for a given formula into a Gaussian distribution within the latent space. Crucially, PCVAE incorporates explicit physics knowledge, such as mappings between crystal systems, Bravais lattices, and space groups, as well as constraints on lattice parameters, to guide the generation process and improve prediction accuracy.

VAEs have also been tailored for specific material applications and enhanced by integrating other machine learning techniques. MagGen \cite{mal_maggen_2024}, for example, is a conditional VAE designed for the inverse design of rare-earth-free permanent magnets. It utilizes Invertible Real space Crystallographic Representation (IRCR) and conditions its latent space on both formation energy and saturation magnetization. A distinctive feature of MagGen is its use of graph theory to analyze the structure of the property-embedded latent space, guiding more efficient sampling towards promising candidates. More recently, WyCryst \cite{zhu_wycryst_2024} integrates Wyckoff position-based crystal representations with a property-directed VAE (PVAE). This approach explicitly ensures that generated crystals comply with crystallographic symmetry rules, particularly Wyckoff site-symmetries, throughout the generation and subsequent DFT refinement stages.

\subsection{GAN-based Models}
GANs have emerged as a popular deep learning paradigm for materials generation. Early applications of GANs in materials science often focus on generating material compositions rather than full crystal structures. A notable example is MatGAN \cite{dan2020generative}, which uses a Wasserstein GAN \cite{arjovsky2017wasserstein} to generate hypothetical compositions of inorganic materials. MatGAN represents materials as sparse matrices indicating elemental counts and successfully learns implicit chemical rules, such as charge neutrality and electronegativity balance. 

The scope of GAN-based generation has expanded from compositions to complete crystal structures of particular material types. Kim et al. \cite{kim_inverse_2020} introduce ZeoGAN for the inverse design of pure-silica zeolites. Using 3D grids for Si/O positions alongside methane potential-energy maps, ZeoGAN reliably produces zeolites matching a user-chosen methane heat of adsorption. GANCSP \cite{kim_generative_2020} develops a composition-conditioned crystal GAN for ternary Mg–Mn–O phases. An inversion-free point-cloud representation, combined with composition conditioning and data augmentation, allows the model to explore structures beyond known templates. In a related line of work, CubicGAN \cite{zhao2021high} aims to the high-throughput discovery of cubic crystals. By conditioning on compositions and space group, it encodes lattice parameters, fractional coordinates, composition embeddings, and symmetry, generating new DFT-validated prototypes. PGCGM \cite{zhao_physics_2023} further integrates physics-based losses tied to pairwise distances and symmetry. This addition boosts generation validity and diversity across 20 space groups, highlighting the benefit of explicit physical guidance.

Researchers are advancing GAN-based materials generation by developing sophisticated crystal representations, integrating property-driven constraints and optimization techniques, incorporating physical knowledge, and addressing intrinsic GAN limitations. For instance, CCDCGAN \cite{long_constrained_2021} employs autoencoder-compressed voxelized representations and trains GANs with direct energy minimization. DeepCSP \cite{ye2024organic} pairs a conditional GAN with a property-predicting MolGAT to rank organic crystal candidates by density. To tackle complex multi-objective optimization, NSGAN \cite{li_nsgan_2024} synergizes GANs with NSGA-II, enabling efficient search within a GAN-learned latent space.

Other notable advancements also include the explicit incorporation of physical knowledge into the generative process and the development of strategies to enhance training stability. CGWGAN \cite{su2024cgwgan} utilizes a Wyckoff-based GAN with a modular approach—template generation, atom infilling, and screening—to enforce crystallographic symmetry. Furthermore, to mitigate issues like mode collapse, GAN-DDLSF \cite{chen_crystal_2024} proposes a data-driven latent space fusion strategy, blending real data statistics with Gaussian noise to enhance generation diversity and realism.

Collectively, these GAN-based models show a progression in the field, moving from compositional generation to sophisticated, property-driven, and symmetry-constrained crystal structure prediction. The development of novel material representations, advanced conditioning techniques, integration with optimization algorithms, and strategies to mitigate common GAN training issues like mode collapse are key trends driving the success of GANs in accelerating materials discovery.

\subsection{Diffusion-based Models}
Diffusion-based models are widely applied for generating geometric graphs of crystals. Crystal Diffusion Variational Auto-Encoder (CDVAE)~\cite{xie2021crystal} integrates diffusion models within a VAE framework. CDVAE comprises three core components: an encoder that maps the unit cell to a latent vector, a predictor that infers graph-level features from the latent representation, and a decoder that reconstructs the all-atom structure using score matching with Langevin dynamics (SMLD). Both the encoder and decoder are implemented as periodic graph neural networks (PGNNs), with DimeNet++ and GemNet-dT adopted in the original implementation. Given a prior latent vector $\mathbf{z}$, the inference process consists of two stages: first, lattice parameters, number of atoms, and compositional ratios are predicted from $\mathbf{z}$ for initialization; then, the atom types and coordinates are generated through Langevin dynamics. 

Building upon this framework, many subsequent variants of CDVAE are proposed to enhance the performance or broaden its range of applications. SyMat~\cite{luo2023towards} directly predicts the set of atom types from the latent vector, and aggregates the coordinate denoising terms from edge distance scores using the chain rule. Cond-CDVAE~\cite{luo_deep_2024} incorporates the composition and pressure as additional conditions into the latent space for crystal structure prediction. To design structures targeting on specific properties, Con-CDVAE~\cite{ye2024cdvae} introduces a conditioning embedding block and a DDPM-based prior generator to generate latent vectors based on required conditions. DP-CDVAE~\cite{pakornchote_diffusion_2024} extends the framework by additionally encoding the number of atoms via an MLP and geometric features through a graph isomorphism network (GINE), and replaces the coordinate generative model from SMLD with DDPM. P-CDVAE~\cite{kaszuba_compositional_2023} predicts the phase of the structure from the latent space and integrates phase conditioning during decoding to distinguish between FCC and BCC crystals. LCOMs~\cite{qi_latent_2023}  improves the optimization process of CDVAE from gradient descent to latent conservative objective models.

Apart from the two-stage generation pipelines, another line of works adopt \textit{joint} generation paradigms. Generally, the unit cell for generation is decomposed into the lattice matrix $\mathbf{L}$ or its corresponding parameters on continuous space, the fractional coordinates $\mathbf{F}$ on torus space, and the atom types $\mathbf{A}$ on discrete space. Various generative models tailored to each underlying manifold are employed to generate these components jointly. Typically, DiffCSP~\cite{jiao_crystal_2023} utilizes standard DDPM, score matching on wrapped normal distribution, and one-hot diffusion for generating the lattice matrix, fractional coordinates and types, respectively. An EGNN-like backbone model is designed to ensure the rotation equivariance and periodic translation invariance during the generation process. EquiCSP~\cite{lin2024equivariant} builds on DiffCSP by further incorporating lattice permutation equivariance and periodic center-of-mass (CoM) invariance. GemsDiff~\cite{klipfel2024vector} applies GemsNet as the backbone model for the joint diffusion framework to enable equivariant lattice updating, and introduces Frechet Distance with ALIGNN (FAD) for evaluating generative fidelity. FlowMM~\cite{miller2024flowmm} parameterizes the lattice using cell lengths and angles, encodes atom types via binarization, and centers the vector fields of fractional coordinates by subtracting their mean. It learns the conditional vector fields defined over the three manifolds via Riemannian flow matching~\cite{chen2023flow}. CrystalGRW~\cite{tangsongcharoen2025crystalgrw} employs geodesic random walks (GRWs) on these manifolds for generation. 

Training on large-scale crystal datasets, MatterGen~\cite{zeni2025generative} incorporates the joint diffusion framework with the classifier-free guidance (CFG)~\cite{ho2021classifier} method, enabling both high-precision de novo generation and customized generation under single or multiple required conditions. DAO~\cite{wu2025siamese} improves the performance of the crystal structure prediction model by training a siamese energy prediction model to relax the unstable structures for augmenting the training data. 

A recent promising direction involves incorporating space group constraints into joint generative models to produce more physically plausible, symmetry-compliant structures. Given a specific space group and assigned Wyckoff position of each site, DiffCSP++~\cite{jiao2024space} enforces the generation process on the crystal family representations and Wyckoff positions to meet the constraints. SymmCD~\cite{levy2025symmcd} further advances this by enabling discrete generation of site symmetries, removing the requirement for predefined Wyckoff assignments.

Beyond inorganic materials, recent work have also applied diffusion models for generating metal-organic frameworks (MOFs). MOFs typically consist of hundreds of atoms per unit cell, making them challenging to model directly at the atomic level. To address this, a hierarchical approach is ofter adopted, where MOFs are first decomposed into coarse-grained building blocks -- the metal ions and organic linkers, then the all-atom structures are recovered from the block-level views. Similar to inorganic materials, the coarse-grained MOF representation consists of the lattice matrix, the coordinates and identities of the building blocks, and further the orientation of each block. MOFDiff~\cite{fu2023mofdiff} extends the CDVAE framework to MOF generation. After predicting the lattice parameters and number of building blocks from the latent vector, the decoder generates the coordinate and identity of each block via SMLD, where the identity is represented by a continuous embedding vector learned from a contrastive learning approach. The orientations are further determined by an assembly algorithm. MOFFlow~\cite{kim2024mofflow} focuses on a structure prediction task that predicts the final structure given a specific set of building blocks. It proposes a flow matching framework to jointly generate the lattices, coordinates and orientations. 

Other than directly applying diffusion models in data space, several approaches extend them to latent representation spaces, where structures are mapped via pre-defined rules or pre-trained encoders. Uni-Mat~\cite{yang2023scalable} introduces a periodic table based representation as the fractional coordinate of each element, while assigning $(-1,-1,-1)^\top$ to absent elements as a form of "disappearance." SuperDiff~\cite{yuan2024diffusion}  applies DDPM in the composition vector space to discover potential superconductors. StructRepDiff~\cite{sinha_representation-space_2024} encodes crystals as a concatenation of lattice parameters, composition vectors, and expanded EAD representations. To unify the generation of atomic systems like small molecules, inorganic crystal and MOFs, an All-Atom Diffusion Transformer (ADiT)~\cite{joshi2025all} is proposed to employ flow matching in a shared latent space, from which a Transformer-based decoder further reconstructs the all-atom structures.

\subsection{Autoregressive-based Models}
Due to the significant success of autoregressive generation methods in natural language processing and computer vision, many studies have attempted to apply them to the field of materials generation. G-SchNet \cite{gebauer2019symmetry} is a pioneer work of autoregressive-based 3D point generation, which aims to factorize a distribution over point sets that enables autoregressive generation where new points are samples step by step. Although G-SchNet can generate atomic systems, its generation target is not specifically designed for materials and does not account for properties such as periodic invariance, resulting in only moderate performance in crystal De Novo generation \cite{jiao2023crystal}.

xyztransformer \cite{flam2023language} is the first to systematically study the performance of language models in the task of atomic system generation. It demonstrates language models, without any architecture modifications, can generate novel and valid 3D structures by Next-Token Prediction (NTP) training paradigm. It proposes two different strategies, LM-CH and LM-AC, to directly derive sequenced from Crystallographic Information Files (CIFs). By 2023, the explosive popularity of ChatGPT accelerated the adoption of the autoregressive generation paradigm based on large-scale language models. CrystaLLM \cite{antunes2024crystal} leverages LLMs (e.g., GPT-2) trained on millions of CIFs and achieves outstanding performance in crystal structure predictions tasks. Similarly, CrystalLLM \cite{gruver2024fine} performs parameter-efficient fine-tuning on LLaMA-2 using CIFs data and achieves strong performance across multi tasks, such as crystal De Novo generation, and text-conditional materials generation. MatExpert \cite{ding2024matexpert} utilizes LLMs and contrastive learning for the material discovery process. By emulating the workflow of human materials design experts, MatExpert combines retrieval, transition, and generation to identify, modify, and create new materials. Unlike previous methods that train-from-scratch or fine-tune language models, MatLLMSearch \cite{gan2025large} claims that pre-trained LLMs are innate crystal structure generators. It integrates pre-trained LLMs with evolutionary search algorithms, achieving high material generation stability rate.

In the methods mentioned above, the autoregressive generation target sequences are mostly CIFs. However, due to the limitation of CIFs, such as the lack of invariances, many methods have also attempted autoregressively generate other sequence patterns of materials. Unlike natural language and other inherently sequential data, atoms in materials have no intrinsic order, so constructing a unique 1D sequential representation of a material is also a significant challenge. Xiao et al. \cite{xiao2023invertible} develop simplified line-input crystal-encoding system (SLICES), which is a string-based crystal representation that satisfies both invertibility and invariances. Furthermore, they propose a three-step reconstruction scheme based on RNN models from SLICES strings to crystal structures, which is referred to as SLI2Cry. Mat2Seq shares a similar motivation with \cite{xiao2023invertible}, aiming to overcome the challenge that CIFs fail to ensure SE(3) and periodic invariance, it proposes a unique sequence for crystal, thereby provably achieving SE(3) and periodic invariance. Additionally, it follows the approach of CrystaLLM \cite{antunes2024crystal}, training a GPT model to perform autoregressive generation based on proposed sequences. NatureLM \cite{xia2025naturelm} represents materials and their 3D structures as 1D sequences in three steps: flatten the chemical formula, add space group information, and include coordinate information. It is a LLM trained with next-token prediction on a wide range of scientific texts, including natural language, formulas, code, and structured data. Uni-3DAR \cite{lu2025uni} introduces a hierarchical tokenization that compresses sparse 3D structures using an octree and encodes fine-grained details like atom types and spatial coordinates. It further enhances efficiency with two-level subtree compression and masked next-token prediction mechanism for dynamic token positions.

\subsection{Hybrid Models}
Hybrid approaches that integrate multiple generative models have gained traction due to the unique strengths of different models in material generation. Autoregressive large language models (LLMs) excel at modeling discrete values but struggle with continuous values due to their reliance on finite-precision representations. Conversely, diffusion models handle continuous values effectively and ensure equivariance but face challenges with discrete elements. FlowLLM~\cite{sriram2024flowllm} bridges this gap by combining autoregressive LLMs for generating initial material representations with a Riemannian Flow Matching (RFM) model for iterative optimization. This integration enhances the generation of stable, unique, and novel materials.

Autoregressive language models have also been applied to text-based material generation, leveraging their text understanding capabilities. For instance, GenMS~\cite{yang2024generative} combines LLMs and diffusion models to generate materials from natural language descriptions. The LLM converts text into chemical formulas, while the diffusion model generates crystal structures from these formulas. Similarly, TGDMat~\cite{das2025periodic} uses a text-guided diffusion model, where a language model encodes textual knowledge to guide the generation of atomic coordinates, types, and lattice structures for materials.

Autoregressive models have further been used to unify diverse generative tasks. UniGenX~\cite{zhang2025unigenx} transforms sequence and structural generation tasks into a unified next-token prediction framework by integrating autoregressive models with conditional diffusion models. This approach enables the handling of diverse scientific data formats while improving numerical precision through a diffusion head for processing numerical data.

Beyond the previously mentioned approaches, several other hybrid methods have also emerged. LCMGM~\cite{chenebuah_deep_2024} combines VAEs and GANs to generate perovskite materials under geometric constraints, utilizing VAEs for structured latent encoding and GANs for diverse, geometry-aware sampling. VGD-CG~\cite{qin_inverse_2024} further integrates VAEs, GANs, and diffusion models to efficiently design stable, synthesizable semiconductor compositions with target properties.

\subsection{Others}
There are also other methods fall outside the above categories. EMPNN~\cite{klipfel2023equivariant} introduces an equivariant message passing network to learn crystal lattice deformations in an unsupervised manner, enabling consistent generation across different geometric representations. CHGlownet~\cite{nguyen2023hierarchical} employs a hierarchical state space with Generative Flow Networks (GFlowNets) to decompose crystal generation into subtasks, improving efficiency and property control. CrysBFN~\cite{wu2025periodic} uses periodic Bayesian Flow Networks (BFNs) with entropy conditioning and non-monotonic dynamics to better model periodicity in non-Euclidean space, enhancing sampling and generation quality.

\section{Datasets}
\label{sec:datasets}
Open-access repositories such as the Crystallography Open Database (COD) \cite{gravzulis2012crystallography}, which hosts over 520,000 CIF files of published organic and inorganic crystal structures, serve as important sources of crystallographic data. In contrast, subscription-based databases prioritize peer-reviewed quality and curation. For example, the Inorganic Crystal Structure Database (ICSD) \cite{hellenbrandt2004inorganic} includes approximately 318,901 rigorously curated inorganic structures, and the ICDD PDF-5+ \cite{icdd2025} contains more than 1.1 million powder diffraction patterns. Similarly, the Cambridge Structural Database (CSD) \cite{allen2004research} offers over 1.25 million experimentally determined structures. These high-quality datasets support big-data research in pharmaceuticals, functional materials, metal–organic frameworks (MOFs), and more, across both academic and commercial sectors.

Complementing experimental archives are computational platforms that provide high-throughput, first-principles data. The Materials Project \cite{jain2020materials} contains 154,718 DFT-computed inorganic compounds, with comprehensive API access and downloadable CIF files. AFLOW \cite{curtarolo2012aflow} offers over 3.5 million calculated entries, including formation energies, band gaps, and elastic constants, all accessible via a REST/JSON API and CIF exports. Emerging mega-datasets are pushing the boundaries of AI-driven materials discovery. OMat24 \cite{barroso2024open} comprises approximately 118 million single-point DFT calculations, while the Novel Materials Discovery (NOMAD) Laboratory Archive \cite{draxl2019nomad} contains over 12 million raw simulation files, standardized using a unified metadata schema. Intermediate-scale repositories like JARVIS-DFT \cite{choudhary2020joint} (40,000+ structures with multiscale properties) offer a balance between depth and scale. Databases such as OQMD and Alexandria also contribute to this ecosystem: OQMD provides formation energies for tens of thousands of materials based on high-throughput DFT, while Alexandria specializes in molecular thermochemical and quantum mechanical properties. HKUST-CrystDB \cite{caobin2025hkustcrystdb} contributes 718,725 structures derived from a combination of computational and experimental sources, processed using the atomic simulation environment. To enable seamless integration across platforms, these datasets increasingly adopt interoperable formats such as CIF, JSON/API, and ASE-DB, promoting smooth data exchange between experimental and computational workflows.

We compare these datasets across several dimensions: accessibility, size, inclusion of physical properties, data origin (experimental vs. computational), composition type (organic or inorganic), and supported retrieval formats. A summary is provided in Table \ref{table:datasetsummary}.

\begin{table*}[t]
\centering
\caption{
Summaries of existing material datasets. \#Open Access: $\checkmark$ = freely available; $\times$ = subscription or restricted access.
Attribute: $\checkmark$ = contains computed physical properties beyond structures.
E or C: Experimental vs.\ Computational.  \textbf{Note : Data were updated as of April 18, 2025.}
} 
\label{table:datasetsummary}
\resizebox{\textwidth}{!}{%
\begin{tabular}{c|ccccccc}
\toprule
\textbf{Dataset} & \textbf{\#Open Access} & \textbf{\#Structures} & \textbf{Attribute} & \textbf{E or C} & \textbf{In/Organic} & \textbf{Format} & \textbf{Link} \\
\midrule
\textbf{COD}             & $\checkmark$ & 523,874         & $\times$     & Both       & Both       & CIF                          & \href{https://www.crystallography.net/cod/}{COD} \\
\textbf{Materials Project} & $\checkmark$ & 154,718         & $\checkmark$ & C          & Inorganic   & CIF, API                     & \href{https://materialsproject.org/}{Materials Project} \\
\textbf{JARVIS‑DFT}      & $\checkmark$ & 40,000 (3D)\,/\,1,000 (2D) & $\checkmark$ & C & Inorganic & CIF, JSON, API               & \href{https://jarvis.nist.gov/}{JARVIS} \\
\textbf{ICSD}            & $\times$     & 318,901         & $\times$     & E          & Inorganic   & CIF                          & \href{https://icsd.products.fiz-karlsruhe.de/}{ICSD} \\
\textbf{AFLOW}           & $\checkmark$ & 3,530,330       & $\checkmark$ & C          & Inorganic   & API (JSON), CIF        & \href{https://aflow.org/}{AFLOW} \\
\textbf{OQMD}            & $\checkmark$ & 1,226,781       & $\checkmark$ & C          & Inorganic   & JSON, API                    & \href{https://oqmd.org/}{OQMD} \\
\textbf{ICDD (PDF‑5+)}   & $\times$     & 1,104,137       & $\times$     & E          & Both        & PDF, TXT, CIF                & \href{https://www.icdd.com/}{ICDD} \\
\textbf{OMat24}          & $\checkmark$ & 118,000,000     & $\checkmark$ & C          & Inorganic   & ASEDB (LMDB)                 & \href{https://huggingface.co/datasets/facebook/OMAT24}{OMat24} \\
\textbf{HKUST-CrystDB}      & $\checkmark$     & 718,725      & $\checkmark$     & Both          & Inorganic   & ASEDB    & \href{https://huggingface.co/datasets/caobin/HKUST-CrystDB}{HKUST-CrystDB} \\

\textbf{Alexandria}      & $\times$     & 1,500,000+      & $\times$     & C          & Inorganic   & CIF, JSON, DGL, PyG, LMDB    & \href{https://alexandria.icams.rub.de/}{Alexandria} \\
\textbf{CSD}             & $\times$     & 1,250,000+      & $\times$     & E          & Organic     & CIF                          & \href{https://www.ccdc.cam.ac.uk/solutions/about-the-csd/}{CSD} \\
\textbf{NOMAD}           & $\checkmark$ & 19,115,490      & $\checkmark$ & C          & Inorganic   & Raw I/O, Metainfo (JSON)     & \href{https://nomad-lab.eu/}{NOMAD} \\
\bottomrule
\end{tabular}
}

\vspace{0.5em}  %
\begin{minipage}{\textwidth}
\footnotesize
\textbf{Format explanations:}
CIF = Crystallographic Information File; API = Application Programming Interface; JSON = JavaScript Object Notation;  TXT = plain text; 
ASEDB = Atomic Simulation Environment Database format; LMDB = Lightning Memory‐Mapped Database; 
DGL/PyG = Graph neural network formats (Deep Graph Library / PyTorch Geometric); 
Raw I/O = binary structure files; Metainfo = NOMAD metadata schema.
\end{minipage}
\end{table*}

\section{Evaluation Metrics}
\label{sec:evaluation}

\subsection{Distribution Criteria}

Crystal distribution metrics are commonly used to quickly assess the generative performance of different models. Although these metrics are not directly related to physical energy, they provide a fast and effective way to benchmark generative models. We summarize several distribution-related metrics below.

\textbf{Validity}. \textit{Structural validity} is defined as the proportion of generated samples in which the minimum interatomic distance exceeds 0.5~\AA.  
\textit{Compositional validity} is the fraction of samples that satisfy charge neutrality, as verified using the SMACT toolkit \cite{davies2019smact}.

\textbf{Coverage}. The \textit{coverage recall} (\(\mathrm{COV\text{-}R}\)) and \textit{coverage precision} (\(\mathrm{COV\text{-}P}\)) calculate the percentage of the crystals in the testing set and that
in generated samples matched with each other within a fingerprint distance threshold.

\textbf{Property Statistics}. To assess how well the generated samples replicate key material properties, the Wasserstein distance is commonly used to compare the empirical distributions of generated and test-set samples for several quantities:
\begin{equation}
d_{\rho} = W\bigl(p_{\mathrm{gen}}(\rho),\, p_{\mathrm{test}}(\rho)\bigr),
\end{equation}
\begin{equation}
d_{\mathrm{elem}} = W\bigl(p_{\mathrm{gen}}(n_{\mathrm{elem}}),\, p_{\mathrm{test}}(n_{\mathrm{elem}})\bigr).
\end{equation}
\begin{equation}
d_{attr} = W\bigl(p_{\mathrm{gen}}(attr),\, p_{\mathrm{test}}(attr)\bigr),
\end{equation}

Here, \(\rho\) denotes the atomic density, \(n_{\mathrm{elem}}\) is the number of elements in the crystal, and \(attr\) represents its physical attributes, such as formation energy, band gap, and bulk modulus.

\subsection{Energy Criteria}

Validation of a novel structure must be based on energy considerations. As shown in Figure~\ref{fig:crystgen}, a stable crystal structure must meet the following criteria:  
\begin{itemize}[leftmargin=*]
    \item \textbf{Uniqueness} — the structure should not duplicate any known structure.
    \item \textbf{Thermodynamic feasibility} — it should be energetically favorable (related to formation energy).
    \item \textbf{Kinetic stability} — it should remain stable under small perturbations.
\end{itemize}
Satisfying all these conditions indicates that the newly discovered structure is theoretically reasonable.

\textbf{Standardization and Structure Matching}. Initially, the generated structure must be cleaned and standardized. It should pass symmetry validation to ensure the correct distribution of atoms and consistency with its space group. Subsequently, it should be compared against existing crystal structure databases to check for equivalency.  

Experimental crystals may include disordered structures, where atomic sites are partially occupied by different elements. Therefore, structure matching should allow a certain tolerance to avoid misclassifying disordered structures as novel ones.

\textbf{Thermodynamic Stability}. To validate the novelty of a structure, its formation energy must be evaluated. A negative formation energy implies that the formation process is thermodynamically favorable — i.e., the structure releases energy when formed from its elemental constituents. This suggests that the material could exist in nature.

The formation energy is calculated as:
\begin{equation}
E_{\text{form}} = E_{\text{total}} - \sum n_i E_i
\end{equation}
where \(E_{\text{form}}\) is the formation energy, \(E_{\text{total}}\) is the total energy of the structure, \(n_i\) is the number of atoms of the \(i\)-th element, and \(E_i\) is the energy of an atom of element \(i\) in its standard state.

Structures with \(E_{\text{form}} < 0\) are thermodynamically stable. However, to ensure practical synthesizability, the structure should also be compared to the convex hull of known phases. If the structure lies within 0–0.1 eV/atom above the convex hull, it is considered metastable but potentially synthesizable \cite{zeni2025generative}. This threshold helps filter out highly metastable structures that are unlikely to be realized experimentally.

\textbf{Kinetic Stability}. Structures predicted to be experimentally realizable must also be evaluated for kinetic stability. This is typically done by analyzing the phonon spectrum. The absence (or minimal presence) of imaginary phonon frequencies suggests that the structure does not spontaneously distort, indicating a stable atomic arrangement.

Kinetic stability is assessed using density functional perturbation theory, where:
\begin{equation}
\omega(q,s) = \sqrt{\frac{1}{M} \Phi(q,s)}
\end{equation}
Here, \(\omega(q,s)\) is the phonon frequency for wave vector \(q\) and mode \(s\), \(M\) is the atomic mass, and \(\Phi(q,s)\) is the force constant matrix.

A structure that passes standardization, structure matching, and both thermodynamic and kinetic stability tests is considered a novel material candidate \textbf{with potential} for experimental synthesis.

\section{Future Directions and Challenges}
\label{sec:future}

The success of AI for Science (AI4S) has significantly accelerated the pace of materials discovery. However, a vast and largely unexplored space remains in AI-driven materials research. Here, we summarize several promising directions that, to our knowledge, are either challenging to address or have not yet been extensively studied.

\begin{itemize}[leftmargin=*]
  \item \textbf{Doping and Defect Structures}:  
    Low-concentration dopants and structural defects introduce local variations that disrupt the periodicity of ideal crystals. These irregularities pose significant challenges for structural representations in AI models. Nevertheless, many critical material properties and functionalities arise precisely from doping and defects.

  \item \textbf{Material Synthesizability}:  
    Most material-generation studies lack in-depth discussion and analysis of synthesizability. Synthesizability is a complex issue that depends not only on the target structure but also on the synthesis pathway. Evaluating whether AI-generated materials can be synthesized under real-world conditions is crucial for practical applications.

  \item \textbf{Integration with Domain Knowledge}:  
    Material generation efforts aim to reconstruct intrinsic distributions governed by physical laws, such as potential energy surfaces. Integrating detailed chemical and physical domain knowledge—like bonding rules, symmetry constraints, and reaction kinetics—into AI models is essential to improve both prediction accuracy and interpretability.

  \item \textbf{Raw Spectroscopy Data Repositories}:  
    Recent studies show that fusing diverse data modalities from various spectroscopic techniques enhances structural representations. However, this benefit conflicts with the current scarcity of accessible raw spectroscopy data. In crystallography, for example, only selected peaks (e.g., three strongest XRD lines) are typically stored \cite{binsimxrd}, while the majority of raw measurements are discarded. AI4S advancements underscore the need to preserve and share complete raw spectroscopic datasets \cite{hollarek2025opxrd}.

  \item \textbf{AI-Driven Laboratories}:  
    Generative models for materials design should transition into practical tools that drive real-world discovery in AI-integrated laboratories. Such “AI labs” would close the loop between model prediction, automated synthesis, and characterization to accelerate innovation.

  \item \textbf{Cross-Scale Modeling}:  
    Effective material modeling often requires bridging spatial scales from quantum-level interactions to mesoscale structures and macroscopic properties. Developing AI frameworks that seamlessly integrate insights across these scales remains a major challenge and an important frontier.

  \item \textbf{Scalability}:
    Advancing toward scalable AI systems—through the design of larger models and the utilization of larger-scale datasets—is a crucial step in driving progress of problem-solving performance \cite{kaplan2020scaling,hestness2017deep,chen2023uncovering}. However,  this remains largely underexplored in materials science. 

\end{itemize}

\section{Conclusion}
\label{sec:conclusion}

In this paper, we provide a comprehensive review of materials generation in the era of AI. We first provide various representations of materials. We then present mainstream backbone architectures and cutting-edge materials generation approaches. Furthermore, we also introduce common benchmark datasets and evaluation metrics. Finally, we highlight future directions and current challenges. We hope this survey will contribute to the emerging field of AI and materials science by facilitating the application of AI-driven materials discovery methods in real-world.

\appendices
\section{Material categories}

Materials exhibit a wide variety of structural forms, ranging from highly periodic crystalline structures to completely non-periodic systems. Current AI-driven research efforts primarily focus on periodic materials, particularly crystalline solids, which can be broadly categorized based on their dominant bonding type:

\begin{itemize}[leftmargin=*]
\item \textbf{Ionic bonding}: Composed of alternating positively charged cations and negatively charged anions, held together by strong electrostatic forces.
\textit{Examples: Sodium chloride (NaCl), Magnesium oxide (MgO).}

\item \textbf{Metallic bonding}: Consist of metal cations immersed in a "sea" of delocalized valence electrons, resulting in high electrical conductivity and malleability.
\textit{Examples: Copper (Cu), Iron (Fe), Aluminum (Al).}

\item \textbf{Covalent bonding}: Contain atoms connected by covalent bonds, forming extended network structures.
\textit{Examples: Diamond, Silicon carbide (SiC), Quartz (SiO\textsubscript{2}).}

\item \textbf{Molecular bonding}: Built from discrete molecules held together by weak intermolecular interactions, such as Van der Waals forces or hydrogen bonding.
\textit{Examples: Ice (H\textsubscript{2}O), Naphthalene (C\textsubscript{10}H\textsubscript{8}), Solid carbon dioxide (CO\textsubscript{2}).}
\end{itemize}

Beyond crystalline materials, there exists a broad class of non-periodic systems, including quasicrystals and amorphous solids. These materials lack long-range order, making it difficult to define a general representation unit suitable for machine learning models. The challenge becomes even more pronounced when considering complex materials with low-concentration dopants, structural defects, or disorder. Despite their critical importance and wide application in materials science, such non-periodic or defect-containing systems remain significantly underexplored in AI-based research.

\section{CSP in X-ray diffraction}

X-ray diffraction–based structure refinement is a powerful technique for determining crystal structures \cite{binsimxrd}. Since a material’s properties depend critically on its structure, experimental structure determination is vital for all classes of matter—ranging from superconductors to catalysts. A major advance in powder diffraction fitting came with the Rietveld method \cite{rietveld1967line, rietveld1969profile}. Instead of analyzing individual, non‐overlapping Bragg peaks, it fits the entire diffraction pattern using calculated profile and a least‐squares algorithm. By minimizing the difference between observed and calculated profile, the method extracts comprehensive structural information. When the fit converges, the refined parameters in the theoretical model represent the only determinable physical features of the crystal.

Structure refinement involves two fundamental tasks \cite{binsimxrd}. The first is \textbf{in-library identification}, which entails matching measured XRD patterns to known entries in a database. Since 1938, crystallographers have archived known structures in the Powder Diffraction File (PDF). By comparing an unknown pattern with entries in this database, researchers can quickly identify materials. Advances in computing have led to the development of numerous software tools that automate the search--match process. The second task is \textbf{out-of-library identification}, which addresses structures not represented in any existing database. This process begins by determining the crystal symmetry, typically the space group, followed by the refinement of a structural model based on the experimental diffraction data.

Traditionally, crystallographers determine the crystal system based on physical properties such as electrical conductivity, optical characteristics, and thermal behavior. The space group is then inferred by analyzing systematic absences in the diffraction pattern. Subsequently, an approximate chemical formula is proposed, and atoms are assigned to Wyckoff positions. Validation is performed through Rietveld refinement and site-occupancy optimization. This trial-and-error process is time-consuming and highly sensitive to the accuracy of the initial structural guess. Crystal structure prediction can generate highly reliable initial models for a given system. By providing more accurate starting structures, these approaches significantly accelerate the structure determination process.

%



\ifCLASSOPTIONcaptionsoff
  \newpage
\fi



\bibliographystyle{IEEEtran}
\bibliography{ref}
\end{document}